\documentclass{mn2e}
\usepackage{natbib_jrm}
\usepackage{graphicx}
\usepackage{amssymb,times}
\bibpunct
[, ]{(}{)}{;}{a}{}{,}

\def \aj {AJ}
\def \mnras {MNRAS}
\def \pasp {PASP}
\def \apj {ApJ}
\def \apjs {ApJS}
\def \apjl {ApJL}
\def \aap {A\&A}

\def \araa {ARAA}
\def \iaucirc {IAUC}
\newcommand{\kms} {$\mathrm{km\;s^{-1}}$}

\newcommand{\ang} {$\mathrm{\AA\;}$}

\def\lesssim{\mathrel{\hbox{\rlap{\hbox{\lower4pt\hbox{$\sim$}}}\hbox{$<$}}}}
\def\gtrsim{\mathrel{\hbox{\rlap{\hbox{\lower4pt\hbox{$\sim$}}}\hbox{$>$}}}}
\begin{document}
\title[SN2002kg/NGC2403-V37 and SN2003gm]{Faint Supernovae and Supernova Impostors: Case studies of SN 2002kg/NGC2403-V37 and SN 2003gm}
\author[J.R.  Maund et al.]{
\parbox[t]{\textwidth}{\raggedright
J.R. ~Maund$^{1}$\thanks{Email: jrm@astro.as.utexas.edu},
S.J. ~Smartt$^{2}$,
R.-P. ~Kudritzki$^{3}$,
A. ~Pastorello$^{2,4,5}$,
G. ~Nelemans$^{6}$,\\
F. ~Bresolin$^{3}$,
F. ~Patat$^{7}$,
G.F. ~Gilmore$^{8}$ and C.R. ~Benn$^{9}$}
\vspace*{6pt}\\
$^{1}$The University of Texas, McDonald Observatory, 1 University Station, C1402, Austin, Texas 78712-0259, U.S.A.\\
$^{2}$Department of Physics and Astronomy, Queen's University Belfast, Belfast,
BT7 1NN, Northern Ireland, U.K.\\
$^{3}$Institute for Astronomy, University of Hawaii, 2680 Woodlawn Drive, Honolulu, Hawaii 96822, U.S.A.\\
$^{4}$Max-Planck-Institut f\"{u}r Astrophysik, Karl-Schwarzschild-Strasse 1, 85748 Garching, Germany\\
$^{5}$INAF-Osservatorio Astronomico di Padova, Vicolo dell'Osservatorio 5, 35122 Padua, Italy.\\
$^{6}$Department of Astrophysics, IMAPP, Radboud University Nijmegen, PO Box 9010, NL-6500 GL Nijmegen, The Netherlands\\
$^{7}$European Southern Observatory, K Schwarzschild Str. 2, 85748 Garching b. Muenchen, Germany\\
$^{8}$Institute of Astronomy, University of Cambridge, Madingley Road, 
Cambridge, CB3 0HA, U.K.\\
$^{9}$Isaac Newton Group of Telescopes, Apartado 321, Santa Cruz de La 
Palma, E- 38700, Spain\\}
\maketitle
\begin{abstract}
Photometric and spectroscopic observations of the faint Supernovae (SNe) 2002kg
and 2003gm, and their precursors, in NGC 2403 and NGC 5334 respectively, are presented.  The properties of these SNe are discussed in the context of previously proposed scenarios for faint SNe: low mass progenitors producing under-energetic SNe; SNe with ejecta constrained by a circumstellar medium; and outbursts of massive Luminous Blue Variables (LBVs).  The last scenario has been referred to as ``Type V SNe'', ``SN impostors'' or ``fake SNe.''\\
The faint SN 2002kg reached a maximum brightness of $\mathrm{M_{V}=-9.6}$, much fainter than normal type II SNe.  The precursor of SN 2002kg is confirmed to be, as shown in previous work, the LBV NGC2403-V37. Late time photometry of SN 2002kg shows it to be only 0.6 magnitudes fainter at 500 days than at the epoch of discovery. Two spectra of SN 2002kg, with an approximately 1 year interval between observations, show only minor differences.  Strong $\mathrm{Fe II}$ lines are observed in the spectra of SN 2002kg, similar to both the LBV NGC2363-V1 and the type IIn SN 1995G.  The spectrum of SN 2002kg does show strong resolved $\mathrm{[N II]}$ at $\lambda\lambda$6549,6583\ang.  The identified progenitor of SN 2003gm is a bright yellow star, consistent with a F5-G2 supergiant, similar to the identified progenitor of SN 2004et.  SN 2003gm, at the epoch of discovery, was of similar brightness to the possible fake SN 1997bs and the type IIP SNe 1999br and 2005cs.  Photometrically SN 2003gm shows the same decrease in brightness, over the same time period as SN 1997bs.  The light curve and the spectral properties of SN 2003gm are also consistent with some intrinsically faint and low velocity type II SN.  The early time spectra of SN 2003gm are dominated by Balmer emission lines, which at the observed resolution, appear similar to SN 2000ch.  On the basis of the post-discovery photometric and spectroscopic observations presented here we suggest that SN 2003gm is a similar event to SN 1997bs, although the SN/LBV nature of both of these objects is debated.  At 226 days post-discovery the spectrum of SN 2003gm is strongle contaminated by HII region emission lines, and it cannot be confirmed that the precursor star has disappeared.  The presence of strong $\mathrm{[N II]}$ lines, near $\mathrm{H\alpha}$, is suggested as a possible means of identifying objects such as SN 2002kg/NGC2403-V37 as being LBVs - although not as a general classification criterion of all LBVs masquerading as SNe.
\end{abstract}

\section{Introduction}
\label{intro}
The different types of core-collapse supernovae (CCSNe) are produced by the gravitational collapse of the nuclei of massive stars ($\mathrm{M_{initial}>8M_{\odot}}$) at the end of their lives.  Some of these CCSNe are unusually faint and may arise from different types of explosion mechanism and, even, other types of objects misclassified as SNe.  The massive eruptions of Luminous Blue Variables (LBVs) have similarities with some observed faint, low-velocity, interacting SNe \citep{zwick61v}.\\
CCSNe are a heterogeneous class of object demonstrating a wide range of explosion energies and expansion velocities \citep{hamobs03,2003MNRAS.346...97N,2003A&A...404.1077E,pastophd}.  \citet{2005coex.conf..275Z} consider a separate branch of subluminous, low velocity type IIP SNe, such as 1997D, 1999br and 2005cs.  SNe with low explosion energies are considered to be produced by either massive stars forming black holes, which quench the SN explosion, or low mass progenitors giving rise to ONeMg cores and electron capture SNe \citep{heg03}.  Studies of the progenitors of these subluminous SNe favour the latter scenario, however, which have been shown to arise from low mass progenitors \citep{2005astro.ph..1323M,2005astro.ph..7502M}.  Low velocities are also observed in the spectra of type IIn SNe, for which various causes have been proposed.  Observations of SNe 1995G and 1998S show that in some cases the low-velocity spectral features arise from the interaction between the ejecta and a dense circumstellar medium (CSM).  Spectroscopy of some type IIn SNe reveals 2-3 velocity components in their strong Balmer lines.  \citet{past95g} observed, for example, broad, intermediate and narrow components to $\mathrm{H\alpha}$ emission in spectra of 1995G.  Type IIn SNe span a large range in brightness from SN 1997cy, at maximum $M_{V}\approx-20.1$ \citep{2000ApJ...534L..57T}, to SN 2000ch, $M_{V}\approx-12.7$ \citep{wagn00ch}.  The Type IIn SNe 1997cy and 1999E are possibly associated with Gamma Ray Bursts (GRBs 970514 and 980910; \citealt{2000ApJ...534L..57T,2003MNRAS.340..191R}).  The combination of extremely low velocities and faintness has led to the suggestion that some type II SNe are misclassified LBVs undergoing outbursts similar to $\mathrm{\eta}$ Car \citep{zwick61v}.\\
\citet{1994PASP..106.1025H} present a review of LBV characteristics.  A handful of LBVs have been identified in the Galaxy and a few extragalactic LBVs, undergoing outbursts, have also been discovered (e.g. \citealt{2001ApJ...546..484D}).  The LBV phase is an important stage in the evolution of the most massive stars ($\ge 40M_{\odot}$; \citealt{1994PASP..106.1025H}) with irregular episodes of mass loss, including extreme eruption events.  The ejected mass goes on to form a CSM, which may be observed as an LBV nebula (such as for $\eta$ Car).  These massive stars will eventually produce a SN which then interacts with this CSM \citep{1989plbv.coll..135L}.  Such massive stars are also expected to be progenitors of Gamma Ray Bursts when they are in the Wolf-Rayet phase of their evolution \citep{1993ApJ...405..273W}.\\
\citet{zwick61v} identified SN 1961V in NGC 1058 as being peculiarly faint and having similar characteristics to $\eta$ Car; this was the prototype for his type V class of SNe.  A number of other SNe have been noted for their peculiar faintness and low velocities and have been suggested to be LBVs: SN 1954J \citep{2001PASP..113..692S}, 1978K \citep{ryder78k}, 1997bs \citep{vandyk97bs}, 1999bw \citep{1999IAUC.7152....2F}, 2000ch \citep{wagn00ch}, 2001ac \citep{2001IAUC.7597....3M} and 2002kg \citep{weis02kg}.  Some of these objects have been referred to as ``SN impostors'' \citep{vandyk97bs}.  In the case of SN 1961V it is not completely clear if in fact this object was the outburst of an LBV.  The faint, prolonged and erratic light curve and the low velocities \citep{1995AJ....110.2261F} are consistent with an LBV eruption, but \citet{2004AJ....127.2850C} discuss features of SN 1961V, such as the non-thermal radio spectrum, which are more in keeping with a fading SN remnant.  SN 1986J was suggested by \citet{1991AJ....101.1275U} to be the eruption of an $\eta$ Car-like variable, although late time observations showed it to be similar to proper SNe \citep{1991ApJ...372..531L}.  The decrease in brightness and the disappearance of the identified progenitor star of SN 1997bs \citep{2002PASP..114..403L} led \citet{smartt01du} to suggest that SN 1997bs might be a sub-luminous type IIn SN.\\
Studies of CCSNe have been enhanced by the identification of progenitors in pre-explosion imaging.  In the case of the LBVs pre-discovery imaging permits a study of the object at its minimum level of activity.  Some have been shown to be coincident with previously identified variable stars, such as SN 1954J = NGC 2403-V12 \citep{2001PASP..113..692S} and SN 2002kg = NGC 2403-V37 \citep{weis02kg}.  It is not currently clear, however, where the boundary between outbursting LBVs and intrinsically faint type IIn SN falls.  The study of two objects designated SNe 2002kg and 2003gm, and their precursors, is presented here.  Details of these SNe and their host galaxies are presented in Table \ref{tab:snlbv:galtab}.\\

\begin{table*}
\begin{minipage}{180mm}
\caption{\label{tab:snlbv:galtab}Details of the objects studied}
\begin{tabular}{lcccccc}
\hline\hline
Object &   $\mathrm{\alpha_{2000}}$ & $\mathrm{\delta_{2000}}$ & Host Galaxy & $v_{\mathrm{Vir}}$  & Distance & $E(B-V)_{\mathrm{tot}}$  \\
       &                            &                          &             &  ($\mathrm{km\;s^{-1}}$)      &    (m-M) &                \\
\hline
SN 2002kg & $\mathrm{7^{h}37^{m}01^{s}.83}$ & $\mathrm{+65\degr34\arcmin29\arcsec.3}$ & NGC 2403   & 360  & 27.56   & 0.16 \\
SN 2003gm & $\mathrm{13^{h}52^{m}51^{s}.72}$& $\mathrm{-01\degr06\arcmin39\arcsec.2}$ & NGC 5334 & 1433 & 31.56$^{1}$ &0.10\\ 
\hline\hline 
\end{tabular}
\\
{\em $\mathrm{1}$} Kinematical distance modulus quoted by HyperLEDA ($\mathrm{http://www-obs.univ-lyon1.fr/hypercat/}$), assuming $\mathrm{H_{0}=70kms^{-1}Mpc^{-1}}$.\\
{\em $v_{\mathrm{Vir}}$} Recessional velocity, quoted by HyperLEDA, corrected for Local Group in-fall towards the Virgo cluster.\\
$E(B-V)_{\mathrm{tot}}$, Galactic and internal reddenings quoted by HyperLEDA.
\end{minipage}
\end{table*}

\noindent
SN 2002kg was discovered by \citet{02kgiauc1}, brightening over the period from 2002 October 26 (JD2452573.5) to 2004 January 1 (JD2453006).  SN 2002kg took place in the SABc galaxy of NGC 2403, the distance modulus to which has been limited by Cepheid measurements to $\le 27.75$ \citep{2000ApJS..128..431F}.  Early time spectra of 2002kg \citep{02kgiauc1} showed narrow Balmer emission lines, which were unresolved with FWHM $\la 500$\kms.  The peak absolute V-band magnitude was $\sim - 9 $, about 7 magnitudes fainter than measured during the plateau phase of the normal type IIP SN 1999em \citep{2003MNRAS.338..939E}.  \citet{weis02kg} compared HST ACS images with ground-based Isaac Newton Telescope (INT) images which showed that the position of SN 2002kg was coincident with that of the known luminous variable NGC 2403-V37.\\
SN 2003gm was discovered by \citet{03gmiauc1} on 2003 July 6.2 (JD2452826.7), in the SBc galaxy NGC 5334.  The location of SN 2003gm in NGC 5334 is shown as Fig. \ref{fig:snlbv:2003gmpos}.  A spectrum of SN 2003gm, acquired by \citet{03gmiauc2}, showed a featureless continuum dominated by Balmer emission lines.  It was noted that these emission lines were composed of narrow and broad components.   The kinematical distance modulus (see Table \ref{tab:snlbv:galtab}), in conjunction with a moderate extinction, implied the discovery absolute magnitude (after \citealt{03gmiauc1}) $\sim -14.4$, similar to the maximum brightness of the peculiarly faint SN 1999br \citep{past99br}, but faint compared to normal type II SNe \citep{03gmiauc2}.\\  
\begin{figure}
\begin{center}
\caption[V-band image of SN 2003gm in NGC 5334]{V-band image of NGC 5334 from the Asiago 1.82m Copernico Telescope, acquired with the AFOSC camera.  SN 2003gm, 13.66 days after discovery, is indicated by the cross hairs.  On this figure North is up and East is left.}
\label{fig:snlbv:2003gmpos}
\end{center}
\end{figure}
In \S2 the observations of these two objects are discussed, with the results presented in \S3.  These results are discussed in \S4.\\
These objects will be referred to as SN 2002kg/V37 and SN 2003gm throughout this study.  The term ``epoch of discovery'' is used to refer to the date at which the objects were identified and classified as possible SNe.  The term ``precursor,'' therefore, refers to the object identified in pre-discovery images.\\ 
\section{Observations and Data Analysis}
\label{sec:snlbv:obsana}
\subsection{Photometry}
\label{sec:snlbv:obsanaphot}
\noindent
A journal of the photometric observations is presented as Table \ref{tab:snlbv:obsphot}.  
\subsubsection{SN 2002kg/V37}
Observations with the Isaac Newton Telescope (INT) Wide Field Camera (WFC) of SN 2002kg/V37 were available in the archive of the INT Wide Field Survey (WFS)\footnote{http://www.ast.cam.ac.uk/$\sim$wfcsur/index.php}.  The site of SN 2002kg/V37 was observed 1.68 years prior to discovery in the g',r',i', $\mathrm{H\alpha}$ and {\sc [Oiii]} bands and 0.29 years post-discovery in Harris V and I bands.  The WFC is composed of four $4096\times2048$ pixel CCD chips, covering a  $34\arcmin\times34\arcmin$ field of view, with a pixel size of $0.33\arcsec$.  These observations, along with post-discovery observations, were reduced and analysed with the Cambridge Astronomical Survey Unit's (CASU) INT WFS pipeline.   Further analysis of faint objects in these fields, not analysed by the pipeline, was carried out using {\sc Iraf}'s {\sc DAOPhot} package and the {\it allstar} task.  These images were acquired under non-photometric conditions.  B, V and I band photometric calibration was provided, therefore, by comparison with a later epoch of HST ACS imaging.  Linear transformation equations, in particular colours, were calculated for single isolated stars readily identifiable in the two sets of imaging.
\begin{table*}
\begin{minipage}{170mm}
\caption{\label{tab:snlbv:obsphot}Journal of Photometric Observations.}
\begin{tabular}{lllrrrrl}
\\
\multicolumn{8}{c}{{\bf Pre-discovery Observations}}\\
\hline\hline
Object    & Date           & JD        & Phase   &Filters                    &Instrument & Telescope & Observer \\
          &                & (2450000+)& (days)  &                           &           &           &      \\
\hline
{\bf SN 2002kg} & 2001 Feb 20    & 1961.44   &-612.06 & g',  r', i'               & WFC       & INT       & Lennon     \\
          & 2001 Feb 21    & 1962.42   &-611.06 &[OIII],H$\alpha$           & WFC       & INT       & Lennon     \\
{\bf SN 2003gm} & 2001 Aug 29    & 2150.98   &-675.72 & F450W, F606W, F814W       & WFPC2     & HST       & GO-9042$^{\chi}$  \\
\hline\hline
\\
\multicolumn{8}{c}{{\bf Post-discovery Observations}}\\
\hline\hline
Object    & Date           & JD        & Phase   &Filters                    &Instrument & Telescope & Observer \\
          &                & (2450000+)& (days)  &                           &           &           &      \\
\hline
{\bf SN 2002kg}          & 2003 Feb 08    & 2679.49   &105.99 &V, I                      & WFC       & INT       &Gilmore      \\
          & 2004 Aug 17    & 3234.65   &661.15 &F475W, F606W, F658N, F814W & ACS/WFC   & HST       & GO-10182$^{\psi}$\\
          & 2004 Sep 21    & 3269.57   &696.07 &F435W, F625W               & ACS/HRC   & HST       & SNAP-10272$^{\psi}$\\
\\
{\bf SN 2003gm} & 2003 Jul 20    & 2840.36   &13.66& B, R, I                & AFOSC     & CTA       & Pastorello        \\
          & 2004 Feb 04    & 3039.69   &212.99& V, I                &  Aux-port   & WHT       & Benn              \\
          & 2004 May 24    & 3150.10   &323.4 & F435W, F555W, F814W       & ACS/HRC   & HST       & GO-9733$^{\chi}$  \\
\hline\hline
\end{tabular}
\\
{\em INT} = 2.5m Isaac Newton Telescope, La Palma \\
{\em HST} = Hubble Space Telescope\\
{\em CTA} = 1.82m Copernico Telescope, Asiago, Italy\\
{\em WHT} = 4.2m William Herschel Telescope, La Palma\\
{\em $^{\chi}$} PI: S. Smartt\\
{\em $^{\psi}$} PI: A. Filippenko\\
\end{minipage}
\end{table*}
\begin{table*}
\begin{minipage}{170mm}
\caption{\label{tab:snlbv:obsspec}Journal of Spectroscopic Observations.}
\begin{tabular}{llrrrrrrr}
\hline\hline
Object &Date         & JD        & Phase & Range          &  Resolution    & Instrument & Telescope  & Observer \\
       &             & (2450000+)& (days)   &($\mathrm{\AA}$)&($\mathrm{\AA}$)&      &         &      \\
\hline
{\bf SN 2002kg} &2003 Mar 1   & 2699.31  & 125.81 & 3200-7000      & 5.7            & LRIS & Keck    & Smartt \\
       &2004 Feb 18  & 3053.37  & 479.87 & 3200-7210      &2               & LRIS & Keck    & Smartt/Maund \\ 
\\
{\bf SN 2003gm} &2003 Jul 16  & 2837.38  & 10.68 & 3970-7050      &1.3             & TWIN & CA 3.5m & Aguirre/Aceituno\\
       &2003 Aug 07  & 2859.39  & 32.69 & 3800-6830      &0.8-1.7             & ISIS & WHT     & Nelemans/Montgomery \\
       &2004 Feb 18  & 3053.50  & 226.8 & 3360-9510      &10.6            & LRIS & Keck    & Smartt/Maund \\
\hline\hline
\end{tabular}
\\
{\em Keck} = 9.8m Keck I Telescope, Hawaii, U.S.A.\\
{\em CA 3.5m} = 3.5m Telescope, Calar Alto, Spain\\
{\em WHT} = 4.2m William Herschel Telescope, La Palma\\  
\end{minipage}
\end{table*}

\noindent
The post-discovery INT WFC images were photometrically calibrated against the post-discovery HST ACS/WFC images of SN 2002kg/V37.  The proximity of the Harris V and I filters to the standard Johnson-Cousins V and I meant that a colour independent correction was applied.
\subsubsection{SN 2003gm}
Images of SN 2003gm, in B, V, and I, with the 1.82m Copernico Telescope (CTA) Asiago Faint Object Spectrograph and Camera (AFOSC) and the 4.2m William Herschel Telescope (WHT) Aux-port Camera were reduced in the standard manner in {\sc Iraf}, using the package {\it ccdred} \citep{ccdred}.  AFOSC provides a square field of view, with side $8.14\arcmin$ with plate scale $\mathrm{0.47\arcsec\;pix^{-1}}$, and Aux-port has a circular field, with diameter $1.8\arcmin$ and scale $\mathrm{0.11\arcsec\;pix^{-1}}$.  Bias and flatfield corrections were applied with appropriate frames acquired during the observing runs.  Point Spread Function (PSF) photometry was conducted on these frames using the {\sc Iraf} package {\sc DAOPhot}.  The reduced AFOSC frames were photometrically calibrated with stars from the USNO-B1.0 catalogue \citep{monet03}, which fell on the fields of view.  Due to the smaller field of view the Aux-port imaging was photometrically calibrated by comparison with HST WFPC2 F606W and F814W images.  HST WFPC2 imaging of the location of SN 2003gm was acquired as part of program GO-9042 (PI: S. Smartt).  The data gathered for this program were intended to provide pre-explosion imaging in the event of future CCSNe for the identification of the progenitor objects.  The data were retrieved from the HST archive, having been passed through the On-the-fly-recalibration (OTFR) pipeline, and corrected for geometric distortion with the appropriate calibration frames.  Pairs of images were combined using the {\sc Stsdas} task {\it crrej} to remove cosmic rays.   Photometry was conducted with {\sc DAOPhot} with empirically determined aperture corrections and PSFs. The updated photometric zeropoints and charge transfer efficiency corrections of \citet{dolp00cte}\footnote{http://purcell.as.arizona.edu/} were adopted.  Comparison photometry was conducted with {\sc HSTphot} to check for consistency, but was not used for stars in the vicinity of SN 2003gm due to the complexity of this region. Iterations of PSF subtraction were used to isolate and measure the individual stars, in the region of SN 2003gm, which were identified as point sources in the later HST ACS HRC images.  The WFPC2 flight system magnitudes were converted to standard Johnson-Cousins magnitudes using the relations of \citet{holsphot95}, and subsequent updates by \citet{dolphhstphot}.\\
Post-discovery images (F435W, F475W, F606W, F625W, F658N and F814W) of SN 2002kg, acquired with the HST ACS for programs SNAP-10272 and SNAP-10182 (PI: A. Filippenko) with the HRC and WFC respectively, were available in the HST archive.  Three colour imaging (F435W, F555W and F814W) ACS/HRC imaging was acquired of SN 2003gm for program GO-9733 (PI: S. Smartt).  Iterations of PSF photometry, and subtractions, were conducted on the images with the {\sc DAOPhot} package.  Aperture corrections were calculated empirically to $0.5\arcsec$ and subsequently corrected to $\infty$ using the values of \citet{acscoltran}.  The photometry was corrected for charge transfer efficiency, using the equations of \citet{riesscte}, and transformed to Johnson-Cousins magnitudes using the relations of \citet{acscoltran}.\\
Comparisons of photometry between the ground-based, WFPC2 and ACS imaging were complicated by the large differences in plate scale.  In the case of SN 2003gm there is an improvement by a factor of 4 in the plate scale of post-discovery images, acquired with the ACS HRC, over the pre-discovery images, acquired with the WF3 chip of WFPC2.  The photometry of bright stars common to both the WFPC2 and ACS images was consistent.  The post-discovery ACS/HRC images of SN 2003gm were acquired for the purpose of differential astrometry, with pre-explosion images, and for studies of the surrounding environment.\\
The {\sc Iraf} task {\it imarith} was used to subtract narrow band images from appropriate broad-band images (for SN 2002kg/V37 pre-discovery $\mathrm{H\alpha}$-r' and {\sc [Oiii]}-g' and post-discovery ACS/WFC F606W-F658N)\citep{weis02kg}.  A scale factor was applied to the narrow band imaging, to match the fluxes of those observed to be predominantly-continuum emitting sources in the broad band imaging.  The subtraction of the scaled narrow band images removed the continuum emitting sources from the broad band frames, leaving objects with either strong absorption (positive features) or strong emission (negative features) at narrow band wavelengths in the output images.
\subsection{Spectroscopy}
\noindent
Spectroscopy of SN 2002kg/V37 and SN 2003gm was acquired at a number of epochs.  A journal of these observations is presented as Table \ref{tab:snlbv:obsspec}.  The spectra were reduced in the normal manner using the {\sc Iraf} {\sc specred}, {\sc twodspec} and {\sc longslit} packages.  Flatfield and bias corrections were applied to the science frames, which were subsequently wavelength and flux calibrated using observations of arc lamps and flux standard stars acquired during the observing runs.  The spectra were optimally extracted to the 10\% flux level, perpendicular to the direction of dispersion, and the sky background was subtracted.  The FWHM of lines in the spectra of the arc lamps were used to estimate the spectral resolution.
\subsubsection{SN 2002kg/V37}
Spectra of SN 2002kg were acquired at two epochs (0.34 and 1.31 years after discovery respectively) with the Keck I Telescope and the Low Resolution Imaging Spectrometer (LRIS).  The first observation with LRIS, on 2003 Mar 1, used the 300/5000 grating, with the $0.7\arcsec$ slit and only the blue arm to acquire the spectrum.  The spectrum was acquired in a single 1800s exposure, covering the wavelength range of 3200-7000\ang.  The second observation, of 2004 February 18 (0.97 years later), used both the blue and red arms of LRIS with the 600/4000 grating on the blue side and 1200/7500 grating on the red side.  This provided better spectral resolution and better wavelength coverage than the previous Keck observation.  Spectra from three 1800s exposures of SN 2002kg were extracted separately, and flux and wavelength calibrated, before being combined to produce a master spectrum.  The $0.7\arcsec$ slit was again used for this observation.  The same slit position angle was used for both Keck observations of SN 2002kg/V37, and is shown on Fig. \ref{fig:snlbv:02kgdiff}c.   This slit orientation also permitted the observation of three nearby bright objects, which are identified on Fig. \ref{fig:snlbv:02kgdiff}c as objects A, B and C, and spectra.  Spectra of these objects were extracted and the ACS/HRC images show that they are not single stars but composite objects.
\subsubsection{SN 2003gm}
Spectra of SN 2003gm were acquired at two early epochs.  The first observation, acquired by Aguirre and Aceituno, on 2003 July 16, used the TWIN spectrograph of the 3.5m Telescope at Calar Alto.  The blue and red arms were used to provide a coverage of 3970-5040\ang  and 5960-7050\ang  respectively, with a resolution of 1.3\ang.  This observation, a single exposure of 1800s, was used by \citet{03gmiauc2} to provide the initial classification of SN 2003gm.\\
SN 2003gm was observed with the WHT ISIS spectrograph, on 2003 August 7.  The blue and red arms of the ISIS spectrograph were used, with the R600B and R1200R gratings respectively, to provide wavelength coverage of 3800-5480\ang  and 6140-6830 \ang respectively.  The resolutions achieved for the spectra of the blue and red arms were 1.7\ang  and 0.8\ang  respectively.  A late time spectrum of SN 2003gm was acquired, at the same epoch as the second SN 2002kg/V37 observation, with Keck LRIS.  Given the faintness of 2003gm at the time of the observation (0.62 years after discovery) a low spectral resolution setting was used with the 300/5000 grating for both the blue and red arms of LRIS.  SN 2003gm was observed in 7 separate exposures, the spectra from which were extracted and calibrated separately before being combined.  The total exposure time of the separate observations was 10800s.\\
The reduced and extracted spectra of SN2002kg/V37 and SN 2003gm were analysed with the {\sc Starlink} programme {\sc Dipso}.  The spectra were velocity corrected, for the Earth's heliocentric velocity at the time of observation and the recessional velocities of the host galaxies.  The {\it Emission Line Fitting} (ELF) software, within {\sc Dipso}, was used to measure the position, widths and fluxes of spectral lines.  
\subsection{Differential Astrometry and Progenitor Identification}
\noindent
The {\sc Iraf} task {\it geomap} was used to calculate the transformation 
between pre- and post-discovery images using the coordinates of stars  common to both frames.  The uncertainty in the positions calculated between frames was calculated as the sum in quadrature of the total uncertainty in the transformation and the scatter in the positions of the target estimated from the different centring algorithms used in {\sc DAOphot} photometry.  The positional scatter only becomes significant for estimates of the centres of subsampled PSFs, and hence only for the HST imaging used in this study, as discussed by \citet{2005astro.ph..1323M}.

%
%
%
%
%
\section{Observational Results}
\label{sec:snlbv:results}
\noindent
\subsection{Photometric Results}
\label{sec:snlbv:resphot}
\noindent
\subsubsection{SN 2002kg/V37}
\label{sec:snlbv:2002kgphot}
\noindent
\citet{weis02kg} identified the precursor object on these pre-discovery images, utilising an absolute astrometric technique.  This result is confirmed with the pre- and post-discovery images presented here.  The study presented here utilises the pre- and post-discovery images to compare and contrast the behaviour of SN 2002kg/V37 with LBVs and SNe.  The post-discovery ACS HRC F625W image was aligned with the pre-discovery INT WFC g' image, using 10 stars common to both images.  SN 2002kg/V37 was readily identifiable on the post-discovery INT/WFC V-band and ACS frames.  This permitted the precursor object to be identified on the pre-discovery frame, with an accuracy of 0.25 INT WFC pixels ($0.08\arcsec$).   SN 2002kg/V37 and its precursor object are identified on post- and pre-discovery images on Fig. \ref{fig:snlbv:02kgdiff}.\\
\begin{figure*}
\caption[Pre- and Post-discovery images of the site of SN 2002kg/V37 in
NGC 2403]{Pre- and Post-discovery images of the site of SN 2002kg/V37 in
NGC 2403.  The same orientation is used for all six panels, with North 
up and East to the left.  SN 2002kg/V37 is located at the centre 
of each of the panels .  a) g' filter image acquired with the INT WFC 2001 
Feb 20 (-612.06 days), prior to discovery.  b) Post-discovery ACS/HRC 
F625W 
image, 2004 Sep 21 (+696.07 days), of SN 2002kg/V37, showing its significant 
increase in 
brightness from the pre-discovery epoch (now comparable to the two nearby 
stars at 2,-2 and 2,-4).  c) Post-discovery INT WFC V-band image, 2003 Feb 
08 (+105.99 days), of SN 2002kg/V37.  The dashed line indicates the 
orientation 
of the 
$0.7\arcsec$ slit for both epochs of Keck spectroscopy.  The clusters for 
which spectra were also acquired are indicated by the labelled cross 
hairs. d) Post-discovery, 2004 Aug 17 (+661.15 days), broad band ACS/WFC 
F606W 
image 
showing the complexity of the region around SN 2002kg/V37.  e) 
Post-discovery, 2004 Aug 17 (+661.15 days), ACS/WFC F658N 
($\mathrm{H\alpha}$) 
image, 
showing SN 2002kg/V37 to be a strong $\mathrm{H\alpha}$ emitter.  f) 
Post-discovery, 2004 Sep 21 (+696.07 days), ACS/HRC F625W image of SN 
2002kg/V37 
(showing the central $10\arcsec\times10\arcsec$ section of panel b).} 
\label{fig:snlbv:02kgdiff} 
\end{figure*} 
\begin{figure}
\begin{center}
\rotatebox{-90}{\includegraphics[width=6cm]{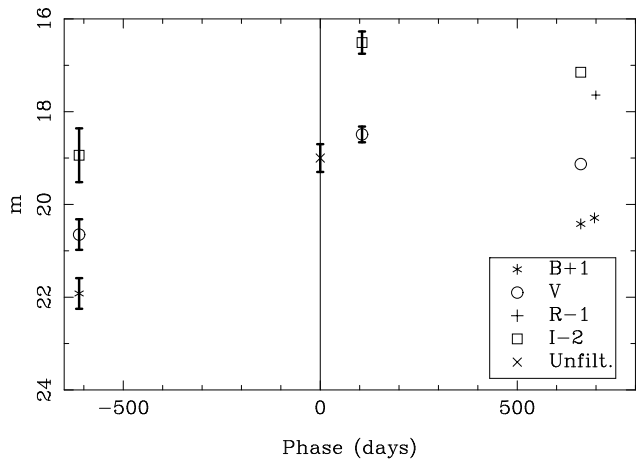}}
\end{center}
\caption[B, V, R and I light curve of SN 2002kg/V37]{The observed B, V, R 
and I light curve of SN 2002kg/V37, relative to the photometry of the 
precursor object.  The photometric uncertainties of data after 661 days 
are smaller than size of the plotted points.  The time scale of the 
$x$-axis 
is measured relative to day 0, the epoch of discovery \citep{02kgiauc1}.}
\label{fig:snlbv:2002kglc}
\end{figure}
\noindent
The photometry of this object from pre-discovery INT/WFC and post-discovery INT/WFC, ACS/HRC and ACS/WFC images is presented in Table \ref{tab:snlbv:2002kgphot}.  A light curve for SN 2002kg/V37 and its precursor is shown as Fig. \ref{fig:snlbv:2002kglc}, based on observations presented here.  An authoritative light curve for SN 2002kg as NGC2403-V37, collected from a variety of sources, is presented by \citet{weis02kg}.\\
The $B-V$ and $V-I$ colours of stars within $2\arcsec$ of SN 2002kg/V37, determined from photometry of the ACS/WFC imaging, was used to determine $E(B-V)=0.17\pm0.02$, which corresponds to $E(V-I)=0.29\pm0.03$.\\
The pre-discovery images provide very poorly constrained colours, principally due to the shallowness of g' and i' images, for the precursor of SN 2002kg/V37 of $(B-V)_{0}=0.4\pm0.4$ and $(V-I)_{0}=-0.6\pm0.6$.  Post-discovery the colour of SN 2002kg/V37 was $(V-I)_{0}=-0.3\pm0.3$.  The later post-discovery ACS/WFC images show SN 2002kg/V37 to have $(B-V)_{0}=0.12\pm0.02$.  The change in brightness from the epoch of the precursor observations ($M_{V}\sim -8.2$) to that at the discovery epoch (0.29 yr later; $M_{V}\sim -10.4$) is modest, with $\Delta V=-2.2\pm0.2$.  The change in brightness is within the range of previous photometric variations observed for NGC2403-V37 \citep{weis02kg}.  The light curve of SN 2002kg/V37 (see Fig. \ref{fig:snlbv:2002kglc}) shows that, post-discovery, the object has not faded rapidly.  At 582 days post discovery SN 2002kg/V37 is only 0.6mags fainter than at the epoch of discovery, although it is not known if there was any variability in the intervening period between points at which the light curve was sampled.
The brightness of SN 2002kg/V37 in the ACS/HRC F625W imaging is consistent with a large contribution of flux from $\mathrm{H\alpha}$ emission, as measured 36 days previously with ACS/WFC.  The subtraction of narrow band images from broad band images has been used previously to show that the precursor of SN 2002kg/V37 was a strong $\mathrm{H\alpha}$ emitter \citep{weis02kg}.  The subtraction of an {\sc [Oiii]} image from a broad band g' image, however, shows nothing at the location of SN 2002kg/V37 in pre-discovery images, despite revealing a nearby ionised region and other diffuse {\sc [Oiii]} emission.  A comparison of post-discovery ACS/WFC narrow band F658N and broad band F606W images shows that SN 2002kg/V37 remains a strong $\mathrm{H\alpha}$ emitter.  This is demonstrated by the subtraction of the F658N image from the F606W image, shown as Fig. \ref{fig:snlbv:02kgsub}c.\\  
\begin{figure*}
\caption[Emission line subtracted broad band images of the site of SN 
2002kg/V37 in NGC 2403]{Emission line subtracted broad band images of the 
site of SN 2002kg/V37 in NGC 2403.  Grey regions indicate locations where 
the flux in the emission line filter is mostly continuum emission, black 
regions indicate where the flux in both the broad band and emission line 
filters arises principally from the emission line.  a) Pre-discovery 
r'-H$\alpha$ frame from the INT WFC observations of 2001 Feb 20-21 
(-612.06 and -611.06 days), SN 2002kg/V37, in the centre of the frame, 
appears black - indicative of a large H$\alpha$ emission.  b) g'-{\sc 
[Oiii]} frame from the INT WFC observations of 2001 Feb 20-21 (-612.06 and -611.06 days).  SN 
2002kg/V37 does not appear in this subtracted frame, suggesting negligible 
flux in the {\sc [Oiii] 5007\ang} line prior to its discovery.  c) 
F658N-F606W ACS/WFC image, from 2004 Aug 17 (+661.15 days), showing that 
post-discovery SN 2002kg is 
again a readily identifiable $\mathrm{H\alpha}$ emitter.  This is also evident in the spectra of SN 2002kg/V37 presented in section \ref{sec:snlbv:2002kgspec}.}
\label{fig:snlbv:02kgsub}
\end{figure*}

\begin{table*}
\begin{minipage}{170mm}
\caption{\label{tab:snlbv:2002kgphot}Pre- and post-discovery photometry of SN 2002kg/V37.}
\begin{tabular}{llrrrrrr}
\hline\hline
Date       &JD (+2450000) &Phase$^{a}$ &     B      &       V   &     R      &     I     & Unfiltered\\ 
           &(days)        &            &            &           &            &           &         \\
\hline 
2001 Feb 20&1961.44       & -612.06   &20.92(0.33) &20.65(0.15)&   ...      & 20.94(0.58)& ...\\
\\
2002 Oct 26& 2573.5       & 0          &...         &...        &...         &... & 19.0(0.3)\\
2003 Feb 08& 2679.59      & 105.99     &   ...      &18.49(0.17)&   ...      &  18.51(0.24) &...\\
2004 Aug 17& 3234.65      &661.15     &19.42(0.01) &19.13(0.01)&  ...       &19.15(0.03)&...\\ 
2004 Sep 21& 3269.57      &696.07      &19.29(0.07) &    ...    &18.643(0.04)&    ...   &... \\
\hline\hline
\end{tabular}
\\
{\em $^{a}$} The phase is given in days from the date of the reported 
discovery (JD2452574, \citealt{02kgiauc1}).\\
Bracketed numbers indicate the magnitude uncertainty.\\ 
\end{minipage}
\end{table*}
\begin{table*}
\begin{minipage}{170mm}
\caption{\label{tab:snlbv:2003gmphot} Pre- and post-discovery photometry of SN 2003gm.}
\begin{tabular}{lrrrrrrrr}
\hline\hline
Date      &       & JD (+2450000)&Phase$^{a}$& B         & V         & R         & I       & Unfiltered$^{b}$  \\
          &Source & (days)&       &           &           &           &         &  \\
\hline
2001 Aug 29& WFPC2&2150.98   & -676.7         &  ...      &24.24(0.13)&...        &23.39(0.23) &  ...\\
2003 Jun 18.2 & Amateur & 2808.70 & -19       &...        &...        &...      &...&$>$19 \\
\\
2003 Jul 06&Amateur&2826.7    &    0         & ...       &...        &...        &...           & 17 \\       
2003 Jul 11&Amateur &2832.7    &    6         & ...       &...        &...        &...           & 17 \\
2003 Jul 20&AFOSC& 2840.36   &   13.66       &18.48(0.27)      &...        &17.83(0.07)&18.01(0.20)& ...\\
2003 Jul 28&Amateur& 2849.4    &  22.7        &...        &...        &...        &...           & 17.3\\
2004 Feb 04&Aux-port& 3039.7   &  212.99       &...        &23.45(0.21)&...        &23.93(0.56)& ...\\
2004 May 24&ACS/HRC &3150.10   &  323.4       &26.07(0.09)&25.12(0.07)&...        &24.90(0.08)&...\\
\hline\hline
\end{tabular}
\\
{\em $^{a}$} The phase is given in days from the date of the reported discovery (JD2452827.7, \citealt{03gmiauc1}), with discovery magnitude of $\approx17$.\\
{\em $^{b}$} Unfiltered Magnitudes from the Bright Supernova site\\$\mathrm{http://www.rochesterastronomy.org/sn2003/\#2003gm}$.\\
Bracketed numbers indicate the magnitude uncertainty.\\
\end{minipage}
\end{table*}

\subsubsection{SN 2003gm}
\label{sec:snlbv:2003gmphot}
\noindent
Post-discovery imaging with WHT Aux-port was used to limit the position of the precursor of SN 2003gm on the pre-explosion WFPC2 imaging to within $0.3\arcsec$.  This uncertainty is the combination of the uncertainty in the transformation between the Aux-port and the WFPC2 images and the seeing ($\sigma\approx0.29\arcsec$) of the Aux-port imaging.  The site of SN 2003gm on the pre-discovery image is confused by a number of blended objects.  The ACS post-discovery imaging was able to resolve the individual components of the blended features.  The positions of the resolved stars on the ACS frame were transformed to the coordinates of the pre-discovery WFPC2 image ($\mathrm{\Delta r \approx 0.16}$ WF pixels) and used to remove the PSFs of nearby stars on the WFPC2 image.  SN 2003gm and its precursor are both easily identifiable, within the area constrained, by the drop in the brightness in the $V$ and $I$ bands between the WFPC2 pre-discovery and the ACS post-discovery images ($\Delta m_{I}\approx+1.5$mag).  The precursor of SN 2003gm, identified with an accuracy of $0.016\arcsec$, occurred on the WF3 chip of WFPC2.  The photometry of stars, around SN 2003gm, between the WFPC2 and ACS/HRC imaging is consistent to $<0.3$ mag, of the same order as the magnitude uncertainties of objects on the WFPC2 image.  The pre-discovery WFPC2 F450W image was not deep enough to detect an object at the location of SN 2003gm.  SN 2003gm and its precursor object are identified on Fig. \ref{fig:snlbv:03gmtot}.  PSF fitting analysis with {\sc DAOPhot} and {\sc HSTPhot} of the precursor object showed it to be consistent with a single point source.  Similar PSF fitting analysis with the ACS/HRC post-discovery images also showed SN 2003gm to be singular.
\begin{figure*}
\caption[Pre- and Post-discovery images of the site of SN 2003gm in NGC 
5334]{Pre- and Post-discovery images of the site of SN 2003gm in NGC 5334. 
a) Pre-discovery HST WFPC2 WF3 F606W image of the site of SN 2003gm, 
from 2001 Aug 29 (-675.72),the precursor object is indicated by the 
cross hairs.   The bright 
point at 2,-2 is a hotpixel.  b) Post-discovery HST ACS/HRC image of SN 
2003gm, from 2004 May 24 (+323.4 days), which is clearly separated from 
other components by the 
exquisite resolution of the ACS/HRC compared to WFPC2 WF3.  c) 
Post-discovery Aux-port image of the site of SN 2003gm, from 2004 Feb 04 
(+212.99), showing the 
comparative increase in brightness and the area constrained by ground based images for SN 2003gm.}
\label{fig:snlbv:03gmtot}
\end{figure*}
The $B-V$ and $V-I$ colours of stars within $2\arcsec$ of SN 2003gm, from the photometry of the ACS/HRC imaging, were used to estimate the reddening of $E(B-V)=0.05\pm0.02$ and, hence, $E(V-I)=0.08\pm0.03$.
Photometry of SN 2003gm, and its precursor object, is presented in Table \ref{tab:snlbv:2003gmphot}.  The V-band and I-band light curves of SN 2003gm are presented as Fig. \ref{fig:snlbv:2003gmlc}. 
\begin{figure}
\begin{center}
\rotatebox{-90}{\includegraphics[width=6cm]{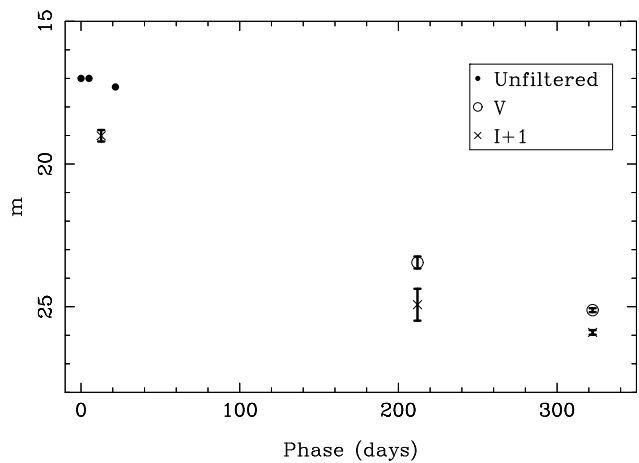}}
\end{center}
\caption[The light curve of SN 2003gm in NGC 5334]{The light curve of SN 2003gm in NGC 5334.  Open squares indicate points of I-band photometry, and open circles show V-band photometry.  Unfiltered reported magnitudes are shown as filled circles.}
\label{fig:snlbv:2003gmlc}
\end{figure}
The precursor object was a star with $(V-I)_{0}=0.8\pm0.3$ and $M_{V}=-7.48\pm0.17$.  The colour is consistent with a mid-F to mid-G yellow supergiant.  The discovery magnitudes of SN 2003gm ($M_{I}=-13.7$) show a large jump in the brightness from the precursor object of $\Delta m_{I}\approx -5.4\pm0.3$.   As presented in Section \ref{intro}, this is substantially fainter than normal type II SNe.  In contrast SN 1995G, a confirmed SN explosion of type IIn, reached a maximum V-band magnitude of $\approx -18.5$ \citep{past95g}.  The paucity of photometric observations of SN 2003gm limits the conclusions one may draw from the light curve about the nature of SN 2003gm and precludes the application of possible classifying criteria (e.g. the presence of a plateau, a linear decline or erratic variability).  The unfiltered photometric points suggest that the early light curve was not rising at the time of discovery.

\subsection{Spectroscopic Results}
\label{sec:snlbv:resspec}

\subsubsection{SN 2002kg/V37}
\label{sec:snlbv:2002kgspec}
\begin{figure*}
\rotatebox{-90}{\includegraphics[width=8cm]{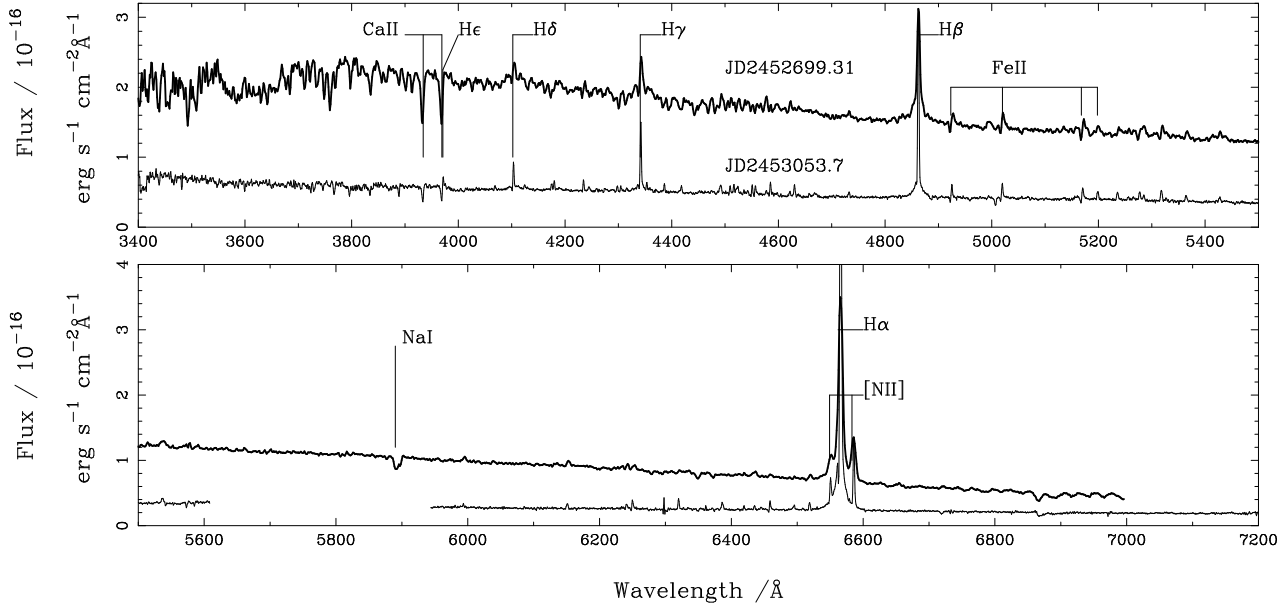}}
\caption[Spectra of SN 2002kg/V37]{Spectra of SN 2002kg/V37. {\it Top 
Panel} Blue flux spectra of SN 2002kg acquired with the Keck LRIS 
instrument.  {\it Bottom Panel} Red flux spectra of SN 2002kg/V37.  These 
spectra have been corrected for the heliocentric velocity at the epochs of 
observation and the recessional velocity of the host galaxy NGC 2403.  The 
spectrum acquired on JD2452699.31 (+125.81 days) is indicated by the heavy 
line and 
the spectrum of JD2453053.37 (+479.87 days) is the thin line.  The break in 
the spectrum 
of JD2453053.7 is due to a gap in the wavelength coverage between the blue and red arms of LRIS.}
\label{fig:snlbv:2002kgfspec}
\end{figure*}

\begin{figure*}
\begin{center}
\rotatebox{-90}{\includegraphics[width=4.5cm]{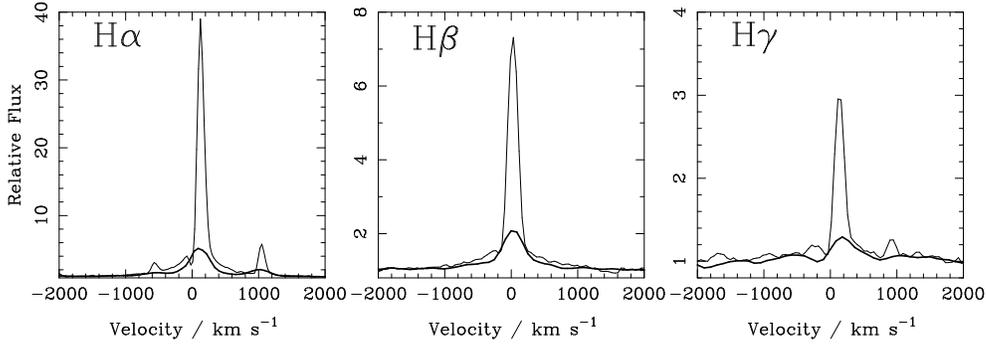}}
\end{center}
\caption[The evolution of the Balmer line profiles of SN 2002kg/V37]{The evolution of the Balmer line profiles in spectra of SN 2002kg/V37.  The spectra from the first and second epoch are shown with the heavy and thin lines respectively.}
\label{fig:snlbv:2002kgwidth}
\end{figure*}

\noindent
 The two spectroscopic observations of SN 2002kg/V37 were conducted with the Keck LRIS instrument, with an interval of 0.97yrs.  The orientation of the $0.7\arcsec$ slit and the clusters which fell along the slit, for the two epochs of observations, are shown on Fig. \ref{fig:snlbv:02kgdiff}c.
Flux spectra from the two epochs of spectroscopic observations are shown as Fig. \ref{fig:snlbv:2002kgfspec}.  Despite the interval between the two observations there are many similarities between the two spectra.  The shape of the continua of the two epochs of spectroscopy mirror the photometric evolution of Fig. \ref{fig:snlbv:2002kglc}, with the slope of the continuum flattened at the later epoch.  Small evolution is perceptible for weak features (although the interpretation is complicated by the different resolutions of the two observations, with a factor three improvement in the resolution of the second epoch observation over that of the first).  These spectra demonstrate, importantly, that broad-band photometry will be influenced by the strong H Balmer lines.\\
The H Balmer lines dominate the spectra at both epochs, being strong relative to the continuum yet narrower at the second epoch (Table 6).  Strong {\sc [Nii]} lines are observed at 6549 and 6583\ang, clearly resolved from nearby $\mathrm{H\alpha}$.  The $\mathrm{H\alpha}$ and $\mathrm{H\beta}$ lines are seen in the first spectrum as pure emission lines, while the higher Balmer lines $\mathrm{H\gamma - H\epsilon}$ are P Cygni profiles (with the absorptions getting progressively stronger and the emission component weaker).  The P Cygni absorptions are unresolved, which for $H\gamma$ implies $\Delta v <  350\mathrm{km\;s^{-1}}$.  The higher Balmer lines are all in absorption.  At the second epoch P Cygni profiles are observed for $\mathrm{H\alpha - H\epsilon}$.  The $\mathrm{H\alpha}$, $\mathrm{H\beta}$ and $\mathrm{H\gamma}$  emission components are themselves composed of narrow and broad velocity components, leading to a narrow peak, with a broad base.   The narrow component of $\mathrm{H\alpha}$ is just above the resolution limit of the spectroscopic observations, whereas the same component of $\mathrm{H\beta}$ is unresolved.  The broad component of the Balmer lines shows a significant decline in width from the first to the second epoch.  The profiles of $\mathrm{H\alpha}$, $\mathrm{H\beta}$ and $\mathrm{H\gamma}$ at the two epochs are shown in Fig. \ref{fig:snlbv:2002kgwidth}.
A large variety of species are observed in emission in both spectra, in addition to H.  The spectrum is very similar to the spectra obtained of the outbursting LBV NGC2363-V1.  The P Cygni profiles immediately redward of $H\beta$ are identified as FeII lines, by comparison with NGC2363-V1.  The identification of these lines as FeII, rather than HeI (e.g. 4923\ang), is preferred as stronger HeI lines (such as that at 4472\ang) are absent.  Lines from HeII, such as 4686\ang are absent from the spectrum of SN2002kg/V37.  Despite seeing prominent {\sc [Nii]} emission, nebular {\sc [Sii]} lines at 6717\ang and 6732\ang are not significantly detected.  {\sc [Oiii]} emission at 4959\ang and 5007\ang is also undetected, with a large absorption appearing at 5009\ang in the interval between the first and second observations.  The Na I D lines were observed in the spectrum acquired on JD2452699.31 and were used to estimate a reddening, with the relationships of \citet{2003fthp.conf..200T}, of $E(B-V)=0.19-0.72$.  This agrees with the low reddening estimated from the photometry of nearby stars presented in section \ref{sec:snlbv:2002kgphot}.

\begin{figure*}
\rotatebox{-90}{\includegraphics[width=8cm]{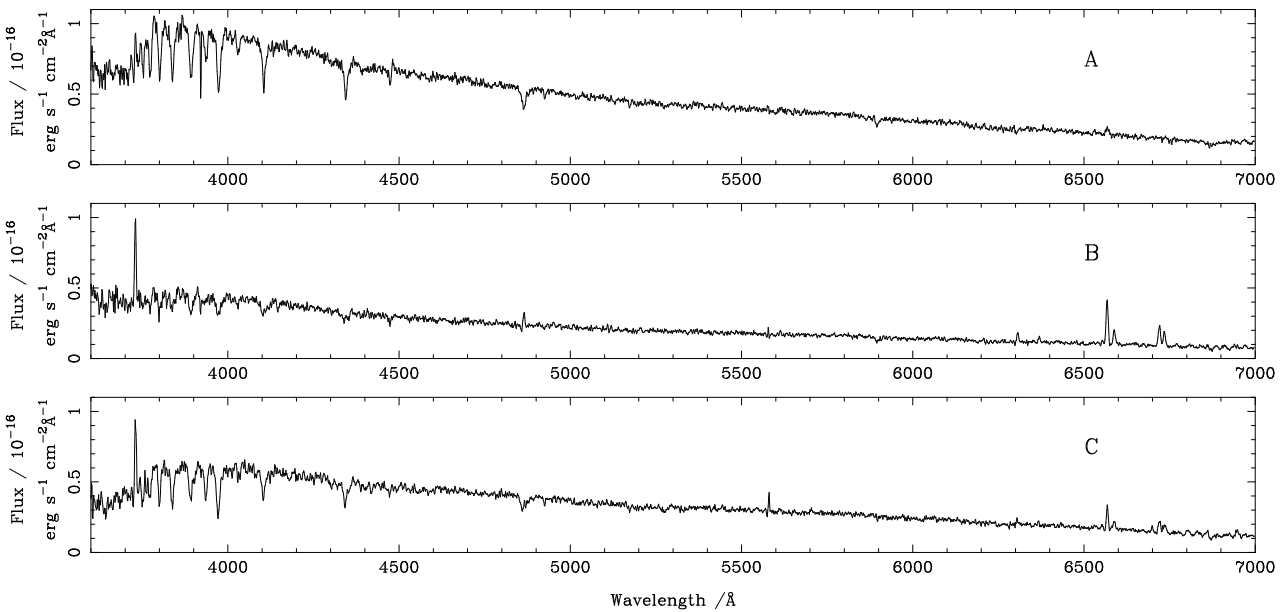}}
\caption[Spectra of three clusters in the vicinity of SN 2002kg/V37]{Spectra of three clusters on the vicinity of SN 2002kg/V37.  The spectra were acquired on 2003 March 1 (JD2453699.31), and these objects are identified on Fig. \ref{fig:snlbv:02kgdiff}c.}
\label{fig:snlbv:clusspec}
\end{figure*}

\begin{table*}
\begin{minipage}{170mm}
\caption[Line strengths in spectra of SN 2002kg/V37]{\label{02kglinestr}Line strengths of features in the spectra, at two epochs, for SN 2002kg/V37.}
\begin{center}
\begin{tabular}{ll|rrrrr}
\hline\hline
                          \multicolumn{2}{r}{JD(+2450000)}&\multicolumn{2}{c}{2699.31}& &\multicolumn{2}{c}{3053.37}\\
\cline{3-4}\cline{6-7}
                                       &           &Flux$^{a}$&FWHM     & &Flux       &FWHM      \\
 Line                                  & Component &          &(\kms)   & & &(\kms)   \\
\hline
$\mathrm{H\alpha}$ 6563\ang            & narrow    &19.5(0.3) &330(3)   & &22.4(0.3) &118(1) \\
                                       & broad     &11.4(0.6)&1570(70) & &12.2(0.6) &790(55)   \\  
$\mathrm{H\beta}$ 4862\ang             & narrow    &8.1(0.3)  &342(9)   & &7.63(0.05)&177(1)\\
                                       & broad     &9.3(0.4)  &1960(150)& &4.4(0.1)  &1360(40)\\
$\mathrm{\left[NII\right]}$ 6549\ang   &           &1.8(0.2)  &unres.   & &1.2(0.2)  &130(30)\\
$\mathrm{\left[NII\right]}$ 6583\ang   &           &4.2(0.2)  &330(11)  & &2.9(0.2)  &114(8)\\ 
\hline
\\
Balmer Dec.$^{b}$                      &           & 1.77(0.06)&     &    &  2.66(0.06)   &       \\
\hline\hline
\end{tabular}
\end{center}
Bracketed numbers indicate the uncertainties of the preceding values.\\
{\em $^{a}$} Flux units $10^{-16}\mathrm{ergs\,s^{-1}\,cm^{-2}\,\AA^{-1}}$.\\
{\em $^{b}$} Ratio of $\mathrm{H\alpha/H\beta}$ of both the narrow and broad emission components, uncorrected for reddening.
\end{minipage}
\end{table*}
\noindent
Objects A, B and C, as identified on Fig. \ref{fig:snlbv:02kgdiff}c, were observed to be non-singular objects in the ACS/WFC post-discovery images.  Spectra of these objects are shown as Fig. \ref{fig:snlbv:clusspec}.  Objects B and C show strong emission line features {\sc [Oii]} and {\sc [Sii]} arising from associated ionised regions, which is linked to the observed nebulosity on the emission line subtracted images on Fig. \ref{fig:snlbv:02kgsub}.

\subsubsection{SN 2003gm}
\label{sec:snlbv:2003gmspec}
\begin{figure*}
\rotatebox{-90}{\includegraphics[width=8cm]{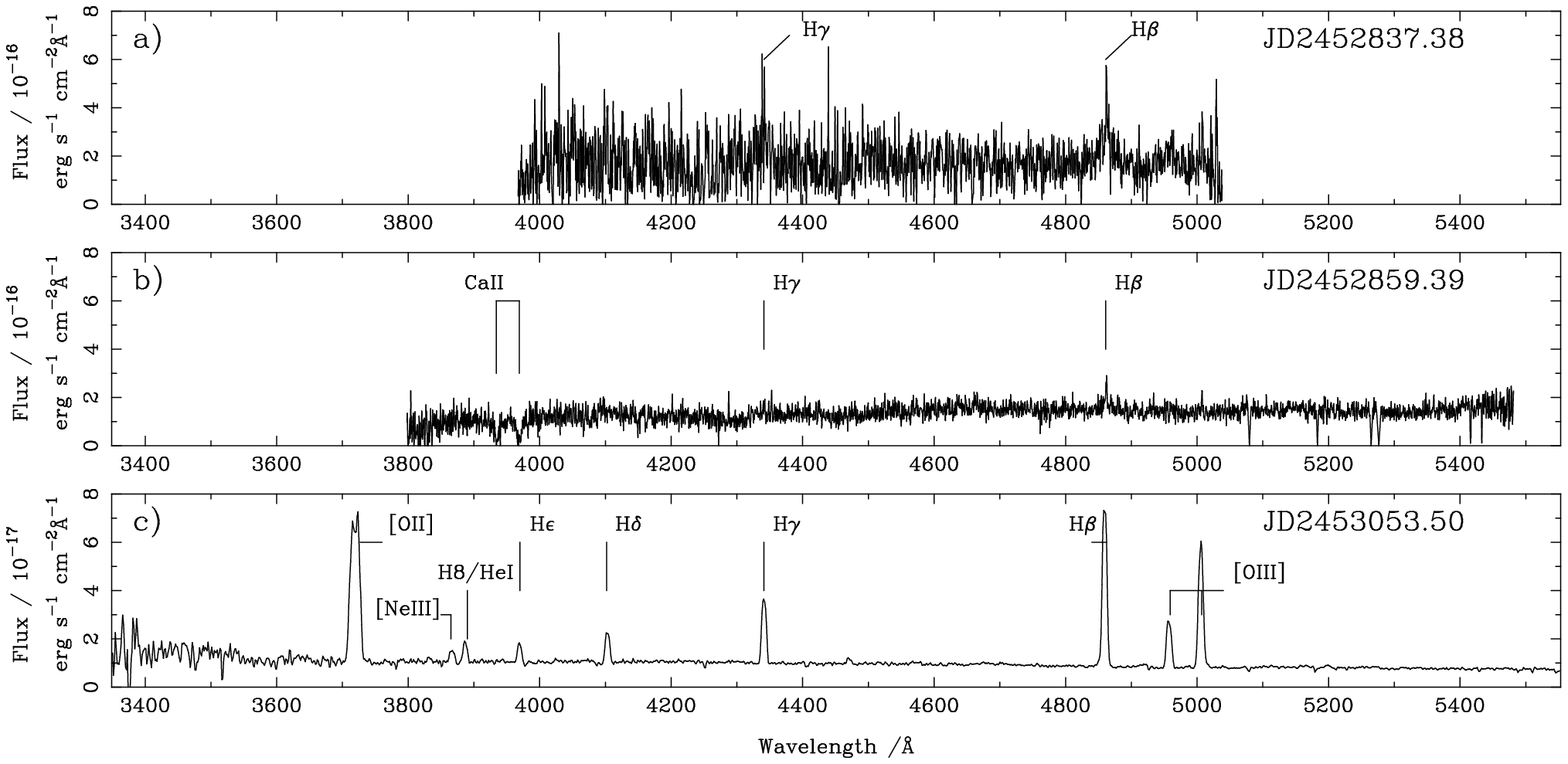}}
\caption[Blue spectra of SN 2003gm in NGC 5334]{Blue spectra of SN 2003gm in NGC 5334 acquired 10.68 days ({\it Panel a)}), 32.69 days ({\it Panel b)}) and 226.8 days ({\it Panel c)}) post-discovery.  The first two epochs show almost featureless continua dominated by strong Balmer emission features.  The latest spectrum shows strong Balmer emission along with ``nebular'' emission features.}
\label{fig:snlbv:2003gmfspecb}
\end{figure*}

\begin{figure*}
\rotatebox{-90}{\includegraphics[width=10cm]{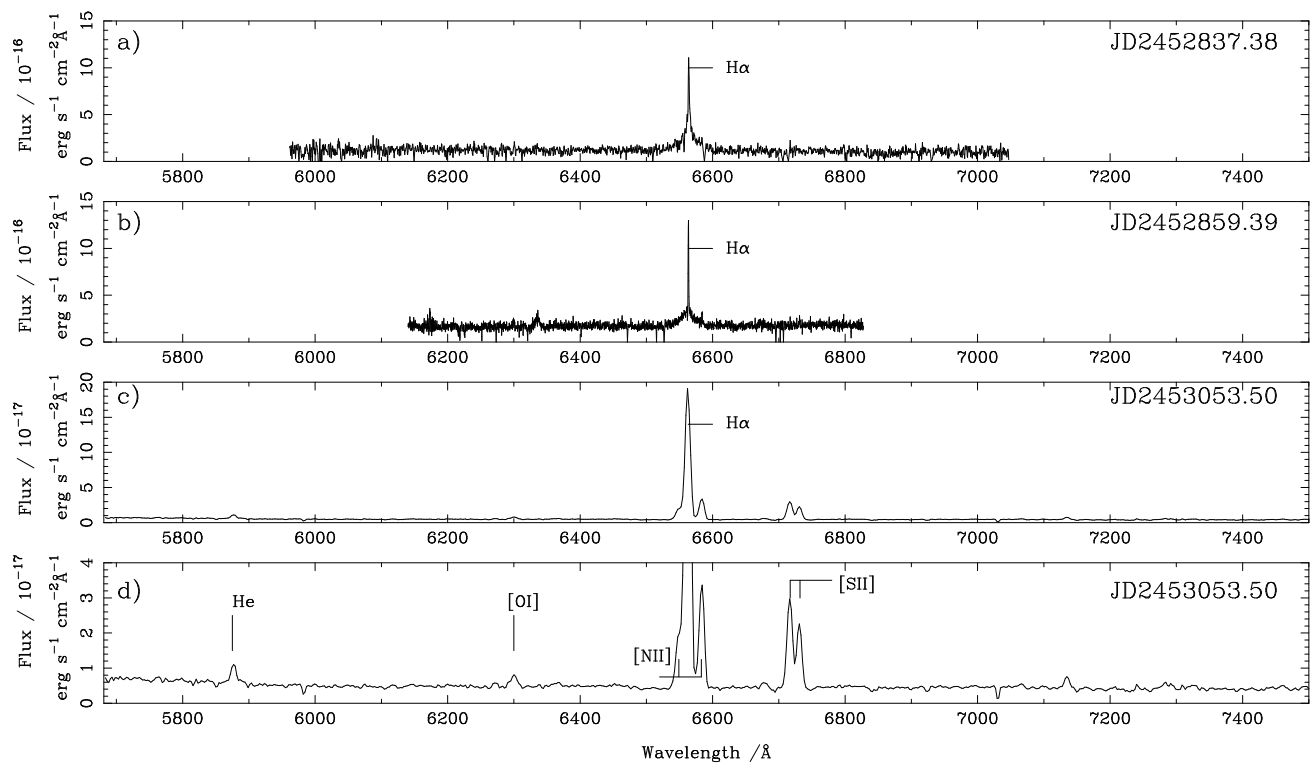}}
\caption[Red spectra of SN 2003gm in NGC 5334]{Red spectra of SN 2003gm in NGC 5334 acquired 10.68 days ({\it Panel a)}), 32.69 days ({\it Panel b)}) and 226.8 days ({\it Panel c) and d)}) post-discovery.  The first two epochs show almost featureless continua dominated by strong Balmer emission features.  The latest spectrum shows strong Balmer emission along with ``nebular'' emission features.  Panel d) is the same as panel c) rescaled to highlight weak features.  Weak emission of HeI (5875\ang), {\sc [O i]} (6300 \ang), {\sc [N ii]} (6549\ang and 6583\ang) and {\sc [S ii]} (6717\ang and 6732\ang) are detected.  These features are more consistent with an HII region.}
\label{fig:snlbv:2003gmfspecr}
\end{figure*}

\noindent
The spectra of SN 2003gm sample a broad time range and show considerable evolution, when compared with the spectra presented in Section \ref{sec:snlbv:2002kgspec}.  The three epochs of spectroscopy are shown as Figs. \ref{fig:snlbv:2003gmfspecb} and \ref{fig:snlbv:2003gmfspecr}.  Measurements of the strengths and velocity widths of certain lines, in these spectra, are presented in Table \ref{tab:snlbv:03gmlinestr}.\\
The two early time spectra, due to the short 22d interval between their acquisitions, are similar.  In both cases the spectra appear as almost featureless continua dominated by Balmer lines.  In both cases, and similarly to SN 2002kg/V37, these Balmer lines are composed of narrow and broad components.  The Balmer lines, in both spectra, are not observed as P Cygni profiles.  Evolution of the spectrum of SN 2003gm, between the two epochs of early time observations, is demonstrated by the decrease in the widths of the Balmer lines.  In contrast to SN 2002kg/V37 the strengths of the Balmer lines, relative to the continuum, have reduced over the interval rather than increased.  Spectra with higher signal-to-noise might have revealed weaker features and the detection of the broad components of higher order Balmer lines.\\

\begin{table*}
\begin{minipage}{170mm}
\caption{\label{tab:snlbv:03gmlinestr}Line strengths of features in the spectra, at three epochs, for SN 2003gm.}
\begin{center}
\begin{tabular}{ll|rrrrrrrr}
\hline\hline
\multicolumn{2}{r}{JD(+2450000)}      &\multicolumn{2}{c}{2837.38}      & &\multicolumn{2}{c}{2859.39}& &\multicolumn{2}{c}{3053.50}\\
\cline{3-4}\cline{6-7}\cline{9-10}
                                      &           &Flux$^{a}$&FWHM      & &Flux$^{a}$&FWHM          & &Flux$^{a}$& FWHM  \\
Line                                  & Component &&(\kms)    & &&(\kms)        & &&(\kms)  \\
\hline
$\mathrm{H\alpha}$ 6563\ang           & narrow    &20.0(1.4) &131(12)   & & 9.8(0.2)& 42(1)   & & 17.3(0.2) & 403(3)\\
                                      & broad     &60.8(3.4) & 1472(120)& &38.1(1.3)& 1170(45)    & &  ...    &  ...   \\  
$\mathrm{H\beta}$ 4862\ang            &narrow     & 5.8(1.0) & 100(18)  & &6.6(0.5) &166(33)        & &5.77(0.05)  &475(5)\\
                                      &broad      &32.2(4.5) &1700(205))& &...      &...     & &   ...   & ... \\
Balmer Dec.                           &           & 2.13(0.28)&         & & 7.3(0.6) &    & & 3.00(0.04) & 
\end{tabular}
\end{center}
Bracketed numbers indicate the uncertainties of the preceding values.\\
{\em $^{a}$} Flux units $10^{-16}\mathrm{ergs\,s^{-1}\,cm^{-2}\,\AA^{-1}}$.\\
{\em $^{b}$} Ratio of $\mathrm{H\alpha/H\beta}$ of both the narrow and broad emission components, uncorrected for reddening.\\
\end{minipage}
\end{table*}

\noindent
The late time Keck spectrum (see Figs. \ref{fig:snlbv:2003gmfspecb} and \ref{fig:snlbv:2003gmfspecr}, acquired 0.62 years after discovery and 0.60 years after the first spectroscopic observation of SN 2003gm, shows a completely different spectrum to the two early time observations.  The spectrum is dominated by the Balmer emission lines, and nebular forbidden emission lines.  The low resolution setting used to acquire this spectrum provides a resolution corresponding to a velocity at $\mathrm{H\alpha}$ of $\mathrm{485km\,s^{-1}}$.  At this resolution none of the observed emission lines are resolved and all are adequately fit by single Gaussian profiles.  The strength of the continuum is greatly reduced compared to the early time spectra, with the forbidden lines growing to comparable strengths as the Balmer lines.  This spectrum is similar to HII regions, such as those presented by \citet{1979MNRAS.189...95P}.  The nebular HeI and {\sc [Oi]} lines observed in late time SN spectra are generally of the same order of strength as $\mathrm{H\alpha}$ \citep{fili97}.  The faintness of SN 2003gm at the epoch of the acquisition of this spectrum is suggestive that this is not a nebular spectrum, but rather an HII region which may be related to SN 2003gm or its precursor.  The spectra of HII regions may be used to determine metallicities and this analysis is discussed in Section \ref{sec:snlbv:disc}.\\
\citet{2004ApJ...615..228B} compare the observed flux of $\mathrm{H\alpha}$ and $\mathrm{H\beta}$ with the intrinsic ratio expected in an HII region.  \citet{1987MNRAS.224..801H} calculate the intrinsic ratio of these line fluxes to be 2.85.  The measured value of this ratio, given in Table \ref{tab:snlbv:03gmlinestr}, was 3.00.  The logarithmic extinction was calculated to be $\mathrm{c(H_{\beta})=0.078}$.  This corresponded to a value of $E(B-V)=0.05$, assuming a \citet{ccm89} reddening law with $R_{V}=3.1$.
The dereddened fluxes of the nebular lines in the spectrum from the latest epoch of observations of SN 2003gm are given in Table \ref{tab:snlbv:hiitab}.
\begin{table}
\caption[Dereddened line fluxes of nebular lines in the late time spectrum of SN 2003gm]{\label{tab:snlbv:hiitab} Dereddened line fluxes of nebular lines in the late time spectrum of SN 2003gm}
\begin{center}
\begin{tabular}{lcc}
\hline\hline
Line             &  Flux$^{a}$         & $\mathrm{c(X)/c(H_{\beta})}^{b}$  \\
\hline
{\sc [Oii]}$\lambda$3727  &$191\pm4$&1.33 \\
$\mathrm{H\delta}\,\lambda\,$4101&$21\pm7$ &1.23\\
$\mathrm{H\gamma}\,\lambda\,$4340&$39.2\pm0.7$ &1.16\\
{\sc [Oiii]} $\lambda$4959&$29.7\pm0.9$ &0.97\\
{\sc [Oiii]} $\lambda$5007&$78\pm1$ &0.97\\
{\sc [Nii]} $\lambda$6548&$38\pm3$ &0.71\\
$\mathrm{H\alpha}\,\lambda\,$6563&$285\pm4$ &0.71\\
{\sc [Nii]} $\lambda$6583 &$47\pm1$&0.71\\
{\sc [Sii]} $\lambda$6717 &$42\pm1$4&0.68\\
{\sc [Sii]} $\lambda$6731 &$27\pm1$&0.68\\
\hline\hline
\end{tabular}
\end{center}
$^{a}$ Fluxes are given in units of $\mathrm{H_{\beta}=100}$.\\
$^{b}$ Assuming a \citet{ccm89} reddening law with $R_{V}=3.1$.\\
\end{table}
%
%
%
%

\section{Discussion}
\label{sec:snlbv:disc}

\subsection{Estimation of the Metallicities of SN 2002kg/V37 and SN 2003gm}
\label{metal}
\noindent
The metallicity of SN 2002kg/V37 was estimated from previous studies of nearby 
HII regions in the host galaxy NGC 2403.  \citet{meta2403} studied HII 
regions in NGC 2403 and modelled the properties of the HII region VS 51, 
within 31$\arcsec$ on the sky of the location of SN 2002kg/V37, determining a value of $12+\mathrm{log}\left(\frac{O}{H}\right)=8.49\pm0.09$.  Another HII region, V44, was studied at a similar radial distance to SN 2002kg/V37 with a value of 
$12+\mathrm{log}\left(\frac{O}{H}\right)=8.53\pm0.10$.  The radial offset 
of SN 2002kg, from the centre of NGC 2403, falls between those of V44 and 
V51 (2.80kpc and 3.48kpc respectively).  The oxygen abundance for SN 2002kg/V37 is taken as the 
mean of these two values: $8.5\pm0.1$.  The same result is obtained by 
applying the $\mathrm{log}\frac{O}{H}$ vs. $\mathrm{M_{B}}$ relationship 
of \citet{metapil}, with the HyperLEDA quoted value of 
$\mathrm{M_{B}=-19.52}$ 
and adopting the abundance gradients of \citet{metapil}.  The solar oxygen abundance of 8.66 \citep{2004A&A...417..751A} implies the metallicity of NGC 2403 at the location of SN 2002kg/V37 is $0.7\pm0.2Z_{\odot}$.\\
An abundance analysis was conducted on the last spectrum of SN 2003gm, 
acquired on 2004 February 18.  The $R_{23}$ and $O3N2$ relationships, of \citet{2004ApJ...615..228B} and \citet{2004MNRAS.348L..59P} respectively, were used to estimate oxygen abundance to be 8.48 and 8.51 using $R_{23}$ and $O3N2$ respectively.  The uncertainties in the fits to the observed spectral lines are much 
smaller than $2\sigma$ uncertainty in the $O3N2$ relation of 
$\pm0.25\mathrm{dex}$.  These values imply a metallicity for SN 2003gm of $\sim0.7Z_{\odot}$.  These results are summarised in Table \ref{metatab}.\\
\begin{table}
\caption[The locations of SN 2002kg/V37 and 2003gm, within their host galaxies, and the metallicities estimated for these objects]{\label{metatab} The locations of SN 2002kg/V37 and 2003gm, within their host galaxies, and the metallicities estimated for these objects.}
\begin{center}
\begin{tabular}{lcrr}
\hline\hline
Object    & Offset (R.A., Decl.)& Radial Offset & Metallicity \\
          &from centre of host  &  (kpc)        & $Z/Z_{\odot}$\\
          &   (arcsec)          &               &  \\
\hline
SN 2002kg & +167.9,-100.2 & 3.12 & 0.7 \\
SN 2003gm & -40.7,11.8 &5.68 & 0.7 \\
\hline\hline
\end{tabular}
\end{center}
\end{table}

\subsection{Emission lines, broad-band photometry and colour-magnitude diagrams}
\label{sec:snlbv:sgparadigm}
\noindent
The pre-discovery broad-band photometry of both objects is affected by large uncertainties. In addition, as shown in section \ref{sec:snlbv:results} (and by \citealt{weis02kg}) SN2002kg/V37 was an unusual emission line object already in pre-discovery phase. Thus, the transformation of colours and magnitudes into luminosity and effective temperature using the colour-effective temperature relationship for normal supergiants \citep{drill00,1999A&A...350..970K} to discuss the physical properties of the objects in the HRD does not seem to be useful. Instead, we compare the location of SN2002kg/V37 and SN2003gm in the colour-magnitude plane with those of similar objects. We will also use the photometry of the surrounding stars to estimate the age of the stellar population. This simpler and very direct approach will provide important constraints on the nature of the objects.


\subsection{SN 2002kg/V37}
\label{sec:snlbv:2002kgdisc}
\begin{figure}
\begin{center}
\rotatebox{-90}{\includegraphics[width=8cm]{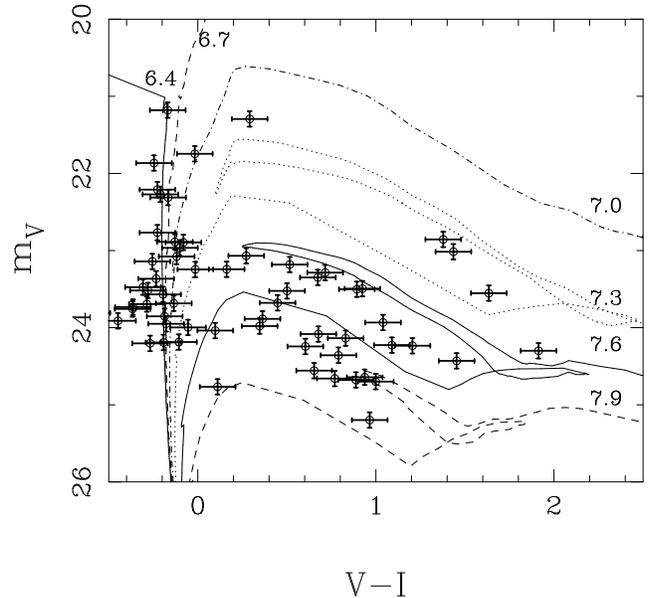}}
\end{center}
\caption[Colour-magnitude diagram showing the positions of stars, in NGC 2403, within $\mathrm{2\arcsec}$ of SN 2002kg/V37]{Colour-magnitude diagram showing the positions of stars, in NGC 2403, within $\mathrm{2\arcsec}$ of SN 2002kg/V37.   Overlaid are half-solar metallicity isochrones shifted for the reddening, extinction and distance to NGC 2403.  A bimodal age distribution is observed with young ($\mathrm{log(age/years)<7.1}$) and old ($\mathrm{log(age/years)=7.6\pm0.1}$) stellar populations.}
\label{fig:snlbv:2002kgcmd}
\end{figure}

\begin{figure*}
\begin{center}
  \rotatebox{-90}{\includegraphics[width=7cm]{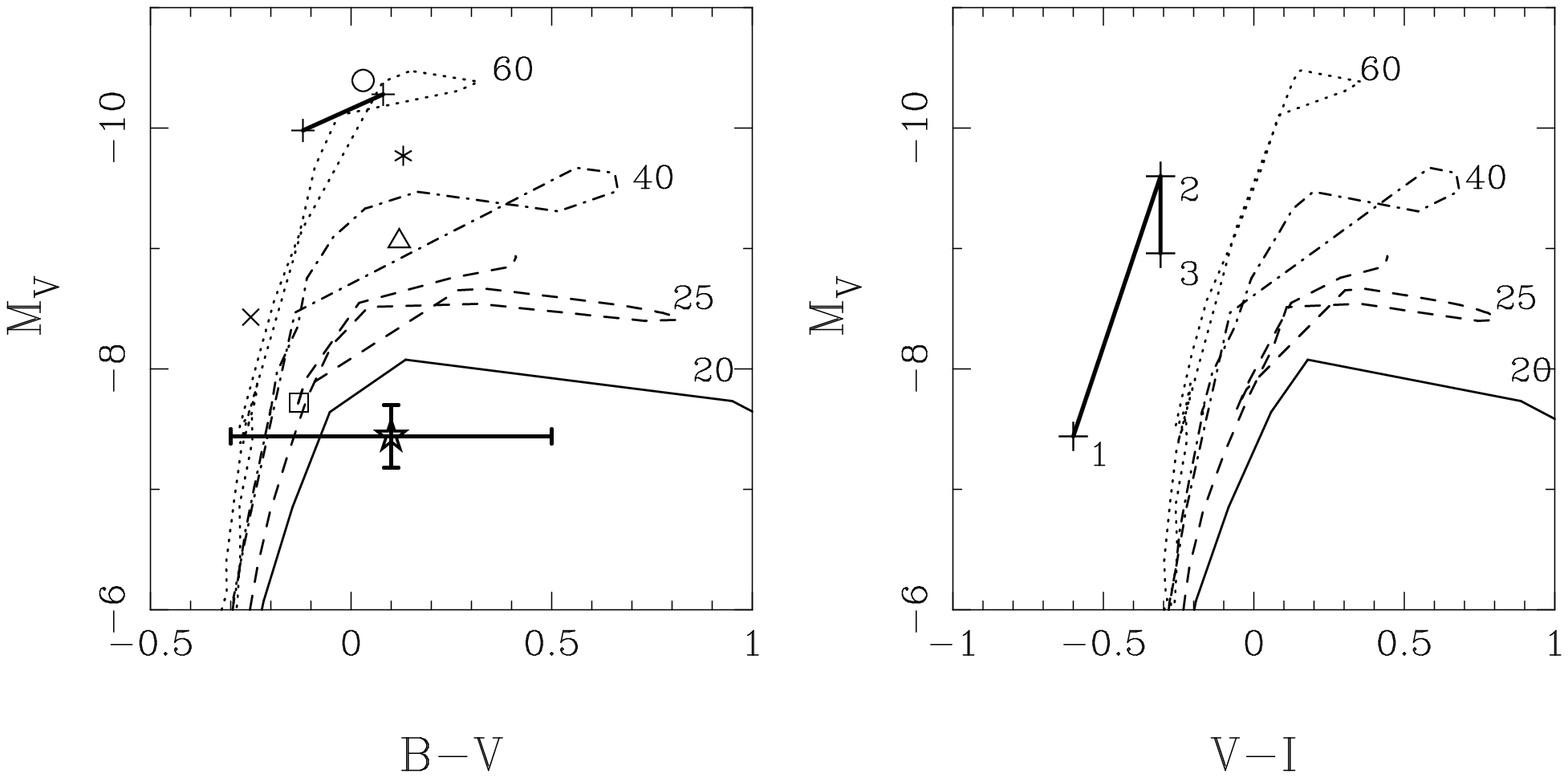}}
\end{center}
  \caption[Colour-magnitude diagrams of the precursor and the post-discovery behaviour of SN 2002kg/V37]{Colour-magnitude diagrams of the precursor and the post-discovery behaviour of SN 2002kg/V37.  {\it Left Panel} The position of the precursor of SN 2002kg/V37 ($\star$).  The positions of six LBVs are also shown on the colour-magnitude diagram: R127($\circ$) and R71 ($\ast$) from \citet{1989A&A...217...87W}, NGC 2363-V1 ($+$, at discovery) from \citet{2001ApJ...546..484D}, P Cygni ($\times$) from \citet{2001A&A...366..935M}, an LBV candidate identified at minimum in NGC 300 ($\vartriangle$) by \citet{2002ApJ...577L.107B} and an LBV candidate identified in NGC 3621 ($\square$) by \citet{2001ApJ...548L.159B}.  {\it Right Panel}  The pre- and post-discovery behaviour of SN 2002kg/V37 are indicated chronologically, following the photometric observations listed in Table \ref{tab:snlbv:2002kgphot}.  Overlaid are half-solar metallicity stellar evolution tracks, with the initial mass (in solar masses) indicated.}
  \label{fig:snlbv:2002kghrd}
\end{figure*}

\noindent
The LBV nature of SN 2002kg/V37 has been previously demonstrated by \citet{weis02kg}.  The results presented here, however, show that post-discovery data alone can be used to determine the LBV nature of such objects.  The spectroscopic study of SN 2002kg provides a comparison for future LBVs misidentified as SNe, which might not be fortunate enough to have been identified as LBVs prior to discovery of the ``SN.''\\
The photometry of stars within $2\arcsec$ of SN 2002kg/V37 was used to estimate an age (see Fig. \ref{fig:snlbv:2002kgcmd}), by comparing their positions on a colour-magnitude diagram with theoretical isochrones (from the database of non-rotating stellar evolution models of the Geneva group and associated isochrones \citep{lej01}).
A bimodal age distribution is observed composed of young ($\mathrm{log(age/years)=6.4-7}$) and old ($\mathrm{log(age/years)=7.6-7.9}$) populations.  Stars from these two populations are evenly distributed together over the 12.5 square arcsecond region around SN 2002kg/V37.  The age of the young stellar population is consistent with the lifetimes of massive stars with $\mathrm{M_{ZAMS}>20M_{\odot}}$, with the lowest age permitted by the uncertainties including the 2.5Myr lifetime of $120M_{\odot}$ stars.  The uncertainty in the age of the younger stellar population arises from the proximity of the low age isochrones to each other.  The age of the older population implies a maximum mass of $\mathrm{6M_{\odot}}$ for the component stars.  This implies that NGC 2403 has undergone a recent epoch of star formation, at the same location as previous generations of stars.  A worthwhile study would be to use the acquired spectra, or expanded broad band photometry, of the nearby clusters to estimate an age, in a similar manner to that of \citet{2004ApJ...615L.113M}, to see if these are similar to the age of the young population of ``field'' stars.  If one considers SN 2002kg/V37 to be either an SN or an LBV then the age requirements for both of these phenomena require the precursor to be a member of the younger stellar population.\\
The brightness and colour of the SN 2002kg/V37 and its precursor are shown on the colour magnitude diagrams of Fig. \ref{fig:snlbv:2002kghrd}.  The positions of a number of other LBVs, at outburst and during quiescence, are also plotted for comparison on the left-panel of Fig. \ref{fig:snlbv:2002kghrd}.
The uncertainty of the pre- and post-discovery colours does not permit an accurate discussion of the relative change in colours before and after discovery.   The location of the precursor is in the region of the colour-magnitude diagram where LBVs in quiescence, such as P Cygni \citep{2001A&A...366..935M}, are found.  The evidence for strong $\mathrm{H\alpha}$ emission in pre-discovery imaging is indicative that the precursor of SN 2002kg/V37 was not a standard blue supergiant. The change in absolute brightness from the pre-discovery to the post-discovery state is small, of the same order as variability previously observed for NGC2403-V37.  This may be compared with the change of absolute brightness observed for the progenitors of CCSNe such as the peculiarly faint SN 1999br, where $\Delta M_{V}\lesssim-7.6$.  The change in brightness observed for SN 2002kg/V37 between the pre- and post-discovery states is small compared to that of LBVs where, following the scheme of \citet{1994PASP..106.1025H}, such a magnitude difference constitutes a ``large excursion'' rather than an ``eruption.''\\  
SN 2002kg follows the general trend of S Dor variables which are blue during quiescence and proceed to the red upon outburst, as the increase in radius causes a drop in temperature as the outburst occurs at a constant luminosity \citep{1989A&A...217...87W}.  A key test would be to check if the pre-discovery state lies on the S Dor strip and the post-discovery state lies on the constant temperature strip for outbursting LBVs on the HR diagram, between which LBVs move as they change from their quiescent state to outburst \citep{1994PASP..106.1025H}.  A number of well studied LBVs are observed to lie in either one of these strips or vary between them \citep{2004ApJ...615..475S}.  The placement of LBVs on the HR diagram requires an accurate effective temperature, which can only be determined with a model atmosphere analysis.\\
The photometric observations, reported on Fig. \ref{fig:snlbv:2002kglc}, show that SN 2002kg/V37 has maintained its brightness over a period of at least 700 days.  Type IIP SNe may, as a consequence of the retention of the progenitor's massive H envelope, maintain their brightness over a period of $\sim100$ days (e.g. 1999em: \citealt{2003MNRAS.338..939E}; \citealt{hamobs03}), followed by a steep decline following the decay lifetimes of radioactive elements such as $\mathrm{^{56}Ni}$.   The light curve of SN 2002kg is also at odds with the shallow linear decay observed for low velocity, interacting SNe such as 1988Z \citep{1993MNRAS.262..128T} and 1995G \citep{past95g}, which for interacting SNe is due to the conversion of kinetic energy to radiation as a consequence of the interaction of the ejecta with the CSM.  The prolonged period of brightness is, however, similar to the observed light curve of the peculiar SN 1961V, which plateaued for $\mathrm{\sim 3}$ months \citep{zwick61v}.  This is also similar to the light curves reported for a number of LBVs such as $\mathrm{\eta}$ Car \citep{1999PASP..111.1124H}, NGC2403-V12(1954J) \citep{1968ApJ...151..825T,2001PASP..113..692S} and NGC 2343-V1 \citep{2001ApJ...546..484D}.  SN 1997bs showed, however, a significant drop in the brightness over a similar time frame.\\
The spectroscopic observations of SN 2002kg/V37 showed minor evolution 
over the 
interval of 0.97yrs.  This is similar to the long term behaviour of the LBV NGC2363-V1 \citep{2001ApJ...546..484D} but not with normal type II SNe \citep{hamobs03}.  The reduction in the velocity widths of the Balmer lines is much smaller than for the type IIn SN 1995G \citep{past95g}, inconsistent with a coasting phase.  The spectrum of SN 2002kg/V37 is very similar to that of NGC2363-V1 and the spectrum of SN 1997bs \citep{vandyk97bs}, such as the relative line strengths and line widths.  The spectra of SN 2002kg/V37 contain many features, albeit narrower, that are found in the spectrum of SN 1995G.  The strong FeII lines in the spectrum are consistent with those observed in LBVs such as NGC2363-V1 and are sometimes assumed to be a classification criterion for identifying LBV outbursts.  There are, however, a large number of FeII lines are also observed in the spectra of SN 1995G \citep{past95g} and no FeII lines are seen in the ``Fake SN'' SN 2000ch \citep{wagn00ch}.  SN 2002kg/V37 does show, however, strong {\sc [Nii]} lines, either side of $\mathrm{H\alpha}$, which are not observed in the type IIn SN 1995G \citep{past95g}.  These strong lines are indicative of raised nitrogen abundance in an ``LBV nebula,'' formed from ejection of N rich material surrounding the LBV \citep{1998ApJ...503..278S}.  The presence of these lines is important due to the conspicuous absence of other nebular lines, such as those listed in Table \ref{tab:snlbv:hiitab}, implying a raised N abundance.  The strengths of the {\sc [Nii]} lines, relative to $\mathrm{H\alpha}$, reported here are smaller than those observed in the LBV nebulae studied by \citet{1998ApJ...503..278S}.  This is because SN 2002kg/V37 is not-spatially resolved from the nebula and so, the observations reported here, contain flux from the star in addition to the nebula.  Other ``nebular'' lines observed by \citet{1998ApJ...503..278S} in the LBV nebulae of R127, S119 and R143 show the other ``nebular'' lines are $\lesssim 20\%$ the strength of the {\sc [Nii]}.  The strengths of the {\sc [Nii]} lines might be used as a classification criterion for LBVs, which have been originally identified as SNe.  This requires the FWHM of $\mathrm{H\alpha}$ to be $\lesssim 1820$\kms, to resolve the {\sc [Nii]} line at $\lambda6582$\ang, and to have good signal-to-noise.  This FWHM limit is less than half that observed for the broad component of $\mathrm{H\alpha}$ in early spectra of SN 1995G \citep{past95g}.  The photometric and spectroscopic observations of SN 2002kg/V37 show that this object was and is an LBV, and present template observations with which future LBV/SNe candidates can be compared.  The diverse range of phenomena observed for LBVs \citep{1994PASP..106.1025H} means that the more data available for definitively classified objects, the better future classification will be.  Having confirmed the LBV nature of SN2002kg/V37 it seems important to obtain accurate information about effective temperature, luminosity and chemical composition. This can be accomplished through a quantitative spectral analysis of our emission line spectra using spherical extended non-LTE line blanketed model atmospheres including stellar winds. Such work for emission line stars in galaxies beyond the Local Group has been carried out recently by \citet{2002ApJ...577L.107B} in NGC 300 and by \citet{2001ApJ...546..484D} for the LBV NGC2363-V1. This will be the next step of our work.

\subsection{SN 2003gm}
\label{sec:snlbv:2003gmdisc}
\begin{figure}
\begin{center}
\rotatebox{-90}{\includegraphics[width=8cm]{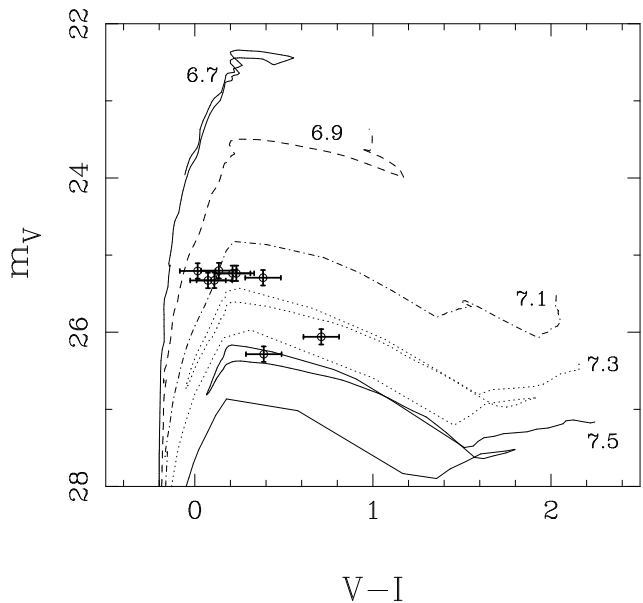}}
\end{center}
\caption[Colour-magnitude diagram showing the positions of stars, from NGC 5334, within $7.5\arcsec$ of SN 2003gm]{Colour-magnitude diagram showing the positions of stars, from NGC 5334, within $7.5\arcsec$ of SN 2003gm.  Overlaid are isochrones, for half-solar metallicity, shifted for the reddening, extinction and distance to NGC 5334.  An approximate age of $\mathrm{log(age/year)}=7.1\pm0.2$ is estimated from this diagram.}
\label{fig:snlbv:2003gmcmd}
\end{figure}

\begin{figure*}
\begin{center}
  \rotatebox{-90}{\includegraphics[width=7cm]{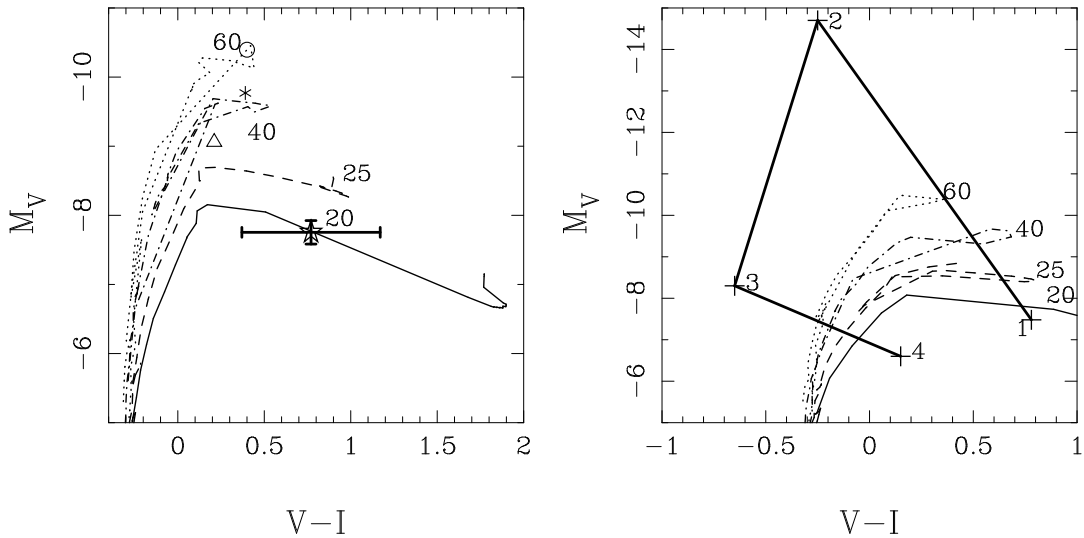}}
\end{center}
  \caption[Colour-magnitude diagrams showing the position of the precursor 
of SN 2003gm]{Colour-magnitude diagrams showing the position of the 
precursor of SN 2003gm ($\star$). 
{\it Left Panel} The positions of three LBVs are also shown on the 
colour-magnitude diagram: R127($\circ$) and R71 ($\ast$) from 
\citet{1989A&A...217...87W} and an LBV candidate identified in NGC 3621 
($\square$) by \citet{2001ApJ...548L.159B}.  {\it Right Panel} The pre- 
and post-discovery behaviour of SN 2003gm are indicated chronologically, 
following the photometric observations listed in Table 
\ref{tab:snlbv:2003gmphot}.  Note that the V-band magnitude at 
discovery, point 2, is assumed to be the same as the reported unfiltered 
magnitude at 0 days, and the corresponding I-band magnitude is that of 
12.7 days.  Overlaid are 
half-solar 
metallicity stellar evolution tracks, with initial masses indicated}
  \label{fig:snlbv:2003gmhrd}
\end{figure*}

\begin{figure}
\begin{center}
  \rotatebox{-90}{\includegraphics[width=8cm]{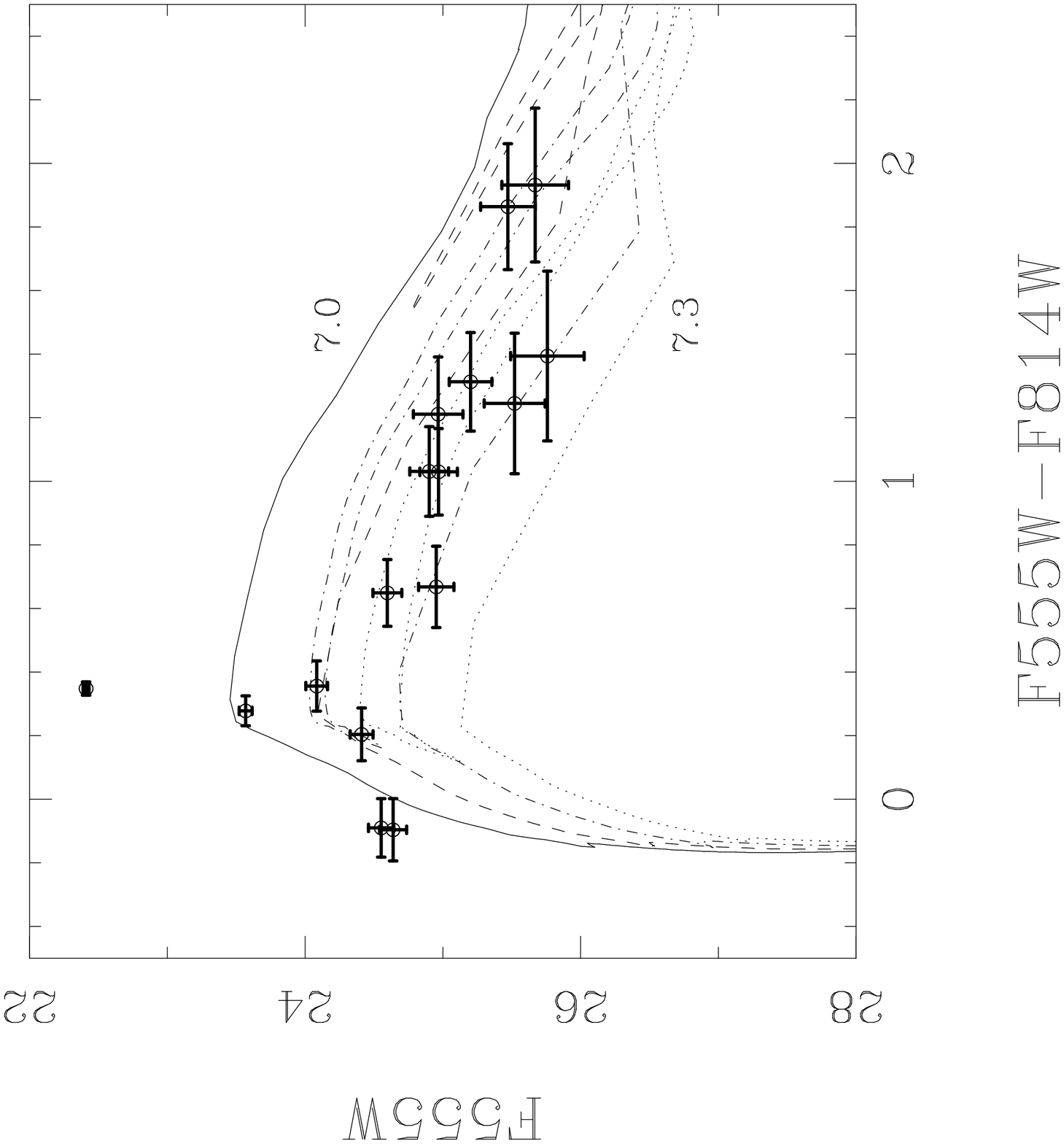}}
\end{center}
  \caption[Colour-magnitude diagram showing the locus of stars within $4\arcsec$ of SN 1997bs in M66]{Colour-magnitude diagram showing the locus of stars within $4\arcsec$ of SN 1997bs in M66.  700s F555W and F814W images from HST WFPC2, from the 2001 March 04 and 2001 February 24 respectively, were photometered separately using HSTphot.  Overlaid are solar metallicity isochrones in the WFPC2 photometric system, for $\mathrm{log(age/years)}$ 7.0 ({\em solid}), 7.1 ({\em dashed}), 7.2 ({\em dot-dashed}) and 7.3 ({\em dotted}).  The isochrones were shifted for the distance, extinction and reddening towards M66, using the values for these parameters given by \citet{vandyk97bs} ($\mu=30.28$, $E(B-V)=0.21$).  \citet{vandyk97bs} presented a colour-magnitude diagram based on the same data as this figure.  A smaller sample area, compared to \citet{vandyk97bs}, has been used (50 sq.$\arcsec$ vs. 240 sq.$\arcsec$), providing an estimate of the age which  is much more appropriate to SN 1997bs.  The age of the surrounding stars is $\mathrm{log(age/years)=7.15\pm0.1}$.  The bright object at the top of the colour-magnitude diagram is a cluster in the vicinity of SN 1997bs.}
  \label{fig:snlbv:1997bscmd}
\end{figure}

\begin{figure}
\begin{center}
  \rotatebox{-90}{\includegraphics[width=6cm]{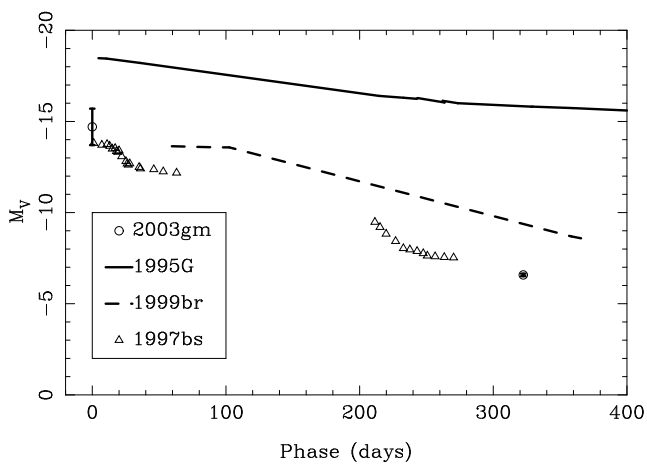}}
\end{center}
  \caption[The late time light curves of SNe 2003gm, 1995G, 1997bs and 1999br]{The late time light curves of SNe 2003gm, 1995G \citep{past95g}, 1997bs \citep{vandyk97bs} and 1999br \citep{past99br}.  The change in V-band brightness of SN 2003gm is similar to that observed for SN 1997bs, covering approximately the same amount of time.}
  \label{fig:snlbv:2003gmcomplc}
\end{figure}

\noindent
The ACS/HRC images were used to provide photometry of stars within 2$\arcsec$ of SN 2003gm (see Fig. \ref{fig:snlbv:2003gmcmd}).  This photometry was plotted on a colour magnitude diagram, with half-solar metallicity isochrones (appropriately shifted for the reddening, extinction and distance to NGC 5334 = 20.5Mpc) to estimate an age for these stars.  An approximate age of $\mathrm{log(age/year)}=7.1\pm0.2$ was estimated from this colour-magnitude diagram, which is similar to that determined for SN 1997bs in M66 (see Fig. \ref{fig:snlbv:1997bscmd}).  This corresponds to the lifetime of stars with initial masses of $\sim25M_{\odot}$.  This is lower than the $40M_{\odot}$ initial mass threshold for stars to undergo an LBV phase in the standard ``Conti'' scenario \citep{1994ARA&A..32..227M}. 
The position of SN 2003gm and its precursor on the colour magnitude diagram are compared with stellar evolution tracks on Fig. \ref{fig:snlbv:2003gmhrd}.
The precursor appears consistent with a yellow supergiant, with $M_{ZAMS}=20M_{\odot}$.  The error bars for the position of the precursor also enclose the points of termination of tracks for stars with $M_{ZAMS}\approx22-23M_{\odot}$.  The stellar evolution tracks of the Geneva group predict that the SN progenitors of stars with $M_{ZAMS}<20M_{\odot}$ should be red supergiants, with stars in the $20-25M_{\odot}$ range being yellower.  In the case of an LBV one might not expect that this colour is representative of the star itself, rather it would be representative of the nebula surrounding the LBV in a similar manner to the Homonculus shrouding $\eta$ Car.  Similarly in the case of SN progenitors circumstellar dust will cause the observed colours to deviate from the actual colours of the progenitor, affecting the inferred spectral type of the progenitor. \citet{2005PASP..117..121L} reported the discovery of a yellow supergiant progenitor for a, seemingly, normal type IIP SN 2004et.  A yellow supergiant was observed as the progenitor of SN 1993J \citep{alder93j}, although this was due to the heavy stripping of the H envelope by a binary companion.  The brightness of the precursor is similar to that of the precursor of SN 1997bs ($M_{V}\approx -7.4$, \citealt{1999AJ....118.2331V}), although there was no colour information for the latter.\\
The light curve for SN 2003gm (see Fig. \ref{fig:snlbv:2003gmlc}) decays over a similar period to which SN 2002kg/V37 was observed to remain at approximately constant brightness.  Caution is required with the interpretation of the light curve of SN 2003gm, which may have varied between observations.  \citet{wagn00ch} showed, however, the erratic light curve of an apparent LBV outburst, SN 2000ch, had a sharp decline after maximum light.  \citet{zwick61v} also showed the light curve of SN 1961V was erratic.   Another LBV candidate, 1997bs, showed a smooth behaviour post-discovery \citep{vandyk97bs}.  This suggests that the shape of the light curve immediately post-maximum is not suitable for identifying LBV outbursts originally classified as SNe.\\ 
The absolute unfiltered magnitude at discovery $M\approx -14$ is similar to the brightnesses for the peculiarly faint SN 1999br \citep{past99br} and the proposed SN impostors SN 1961V \citep{zwick61v} and SN 1997bs \citep{vandyk97bs}, but $\sim1$mag brighter than SN 2000ch at maximum brightness \citep{wagn00ch}.  At maximum brightness this is still $\sim3$mags brighter than NGC2363-V1 \citep{2001ApJ...546..484D}, and corresponds to an increase in brightness from the precursor of $\Delta M_{V}\approx -7$.   The late time light curve of SN 2003gm is compared with the late time light curves of the ``SN impostor'' SN 1997bs, the peculiarly faint SN 1999br and type IIn SN 1995G on Fig. \ref{fig:snlbv:2003gmcomplc}.\\
The two early time spectra of SN 2003gm appear very similar to spectra of SN 2000ch \citep{wagn00ch}, composed of a featureless continuum with large Balmer emission lines.  The host galaxy of SN 2003gm is twice the distance of the host of SN 2000ch and hence the spectra are noisier than those published of SN 2000ch.  Similarly to SN 2000ch {\sc [Nii]} lines are not observed, as might be expected in LBV outbursts, but this may be a consequence of the poor signal to noise rather than the absence of {\sc [Nii]}.  The widths of the Balmer lines are observed to narrow between the first two epochs of spectroscopy.  The fading of SN 2003gm, i.e. the way in which SN features have faded to reveal the spectrum of an HII region, is similar to the way in which SN 1997bs faded \citep{2002PASP..114..403L}.  The manner in which both SN 2003gm and SN 1997bs faded contradicts the suggested rule of \citet{vandyk97bs}: that the precursors of LBVs should remain visible after their outburst.  The spectrum of SN 1997bs, presented by \citet{vandyk97bs}, shows similar Balmer features to SN 2003gm, as well as showing significantly detected FeII lines, which would probably not have been detected in the SN 2003gm spectra due to the levels of noise.\\
The last spectrum shows features consistent with an ionised HII region.  This ionized region could be the host for the progenitor of an SN, ionized prior to explosion, or could be unrelated.  The metallicity determination for SN 2003gm shows no abundance anomalies.   A comparison of line strength ratios from the last spectrum of the site of SN 2003gm with the locus of normal HII regions on the $\mathrm{log([NII]6583\AA/H\alpha)}$ vs. $\mathrm{log([OIII]5007\AA/H\beta)}$ diagram, as presented by \citet{2003MNRAS.346.1055K} as their Fig. 1, shows no unusual line strengths.  The line strength ratios are inconsistent with those observed from the forbidden lines of Seyfert and LINER galaxies, environments which host large shocks that might be expected when SN ejecta collides with the CSM.  The ratio of the {\sc[Nii]} 6583\ang  line strength to that of $\mathrm{H\alpha}$ suggests the ionised region is not an LBV nebula.  In an LBV nebula one expects to see the skewed N abundance due to the mass loss of N rich material as the LBV was undergoing outburst and the elevated levels of mass loss expected for LBVs during minimum.\\
The combination of low luminosity and low velocities suggest that if SN 2003gm is indeed a SN then it is a very under-energetic one.  \citet{zamp99br} predict that progenitors with $M_{ZAMS}\gtrsim 20M_{\odot}$ may form black holes and are, hence, quenched.  SN 2003gm is similar to SN 1997bs, but an authoritative argument for the LBV nature of the latter has not been put forward.

\section{Conclusions}
\label{sec:snlbv:conclu}
\noindent
Post-discovery photometric and spectroscopic observations of two faint, low velocity outbursts have been presented here.\\
The post-discovery observations of SN 2002kg/V37 are consistent with the conclusion of \citet{weis02kg}, who identified SN 2002kg as being the previously identified LBV NGC2403-V37.  The photometric behaviour of SN 2002kg has been shown to be inconsistent with the behaviour of normal interacting type II SN, maintaining its approximate brightness over a period of just under 2 years.  The spectroscopy of SN 2002kg/V37 shows it to be similar to other identified LBVs, such as NGC2363-V1.  Strong {\sc [Nii]} lines were observed in the spectrum, in common with other LBVs, and this feature is suggested for the classification of misidentified LBVs.  The results of this work show that the LBV nature of a SN candidate may be determined over a long period.  If the {\sc [Nii]} lines can be used to identify LBVs, then this work also shows that these features are present in very early spectra and can lead to a classification without the need for observations over an extended period.\\
The nature of SN 2003gm is unclear, although it is photometrically and spectroscopically similar to SN 1997bs .  The photometric behaviour of both of these objects, from the limited observations, could fall within the ``general envelope'' of the behaviour expected for LBVs.  The LBV characteristic features observed in SN 2002kg/V37 were not observed in SN 2003gm.  Future observations of both SN 1997bs and SN 2003gm will show if either of these objects are recurrent variable stars.  High resolution deep imaging of the sites of SNe 1997bs and 2003gm will show whether the precursor object has disappeared, and long term observations, similar to those of NGC2403-V37 \citep{weis02kg}, will show if these objects are recurrent.\\
Progenitor observations have permitted the determination of the relative increase in brightness of the precursor to the objects discovered as SN 2002kg/V37 and SN 2003gm.  In the case of the former the increase in brightness is small.  In the case of SN 2003gm the increase in brightness, while modest, is significantly brighter than most LBVs.\\
Velocity and brightness criteria cannot be used as a definitive classification technique.  SN 2002kg/V37 was of sufficient faintness as to make it extremely unlikely to be a SN.  SN 2003gm was, however of similar brightness to SN 1997bs and only slightly fainter than SN 1999br at the time of discovery.  Strict brightness and velocities limits on the classification scheme would preclude the possible identification of potential sub-luminous type IIn SNe, rather identifying them as LBVs.\\
Spectroscopic classification of these objects with, for example, the {\sc [Nii]} lines is a promising way of distinguishing between the two types of events in the future.  An archive of observations of these objects, for example made available through the {\sc Suspect} website\footnote{$\mathrm{http://bruford.nhn.ou.edu/~suspect/}$}, would permit careful comparison of the properties of such objects discovered and help provide definitive classification criteria in the future.

\section*{Acknowledgments}
JR acknowledges financial support, in the form of a postdoctoral fellowship, from the University of Texas at Austin.  SJS thanks the EURYI scheme for a fellowship and financial support.  Some of data presented here was made publicly available through the Isaac Newton Groups' Wide Field Camera Survey Programme. Some of the data presented here were obtained with the William Herschel Telescope.  The  Isaac Newton Telescope  and William Herschel Telescopes are operated on the island of La Palma by the Isaac Newton Group in the Spanish Observatorio del Roque de los Muchachos of the Instituto de Astrofisica de Canarias.  Some of the data presented herein were obtained at the W.M. Keck Observatory, which is operated as a scientific partnership among the California Institute of Technology, the University of California and the National Aeronautics and Space Administration. The Observatory was made possible by the generous financial support of the W.M. Keck Foundation.  This work uses observations made with the NASA/ESA Hubble Space Telescope, obtained from the data archive at the Space Telescope Science Institute. STScI is operated by the Association of Universitiesfor Research in Astronomy, Inc. under NASA contract NAS 5-26555.  This research has made use of the NASA/IPAC Extragalactic Database (NED) which is operated by the Jet Propulsion Laboratory, California Institute of Technology, under contract with the National Aeronautics and Space Administration.  This research has made use of the HyperLEDA databases ($\mathrm{http://www-obs.univ-lyon1.fr/hypercat/intro.html}$).  Some SNe spectra were retrieved from the SUSPECT Online Supernova Spectrum Archive.  JRM thanks P.A. Crowther, J. Fabbri, M. Irwin, T. Kim, A.D. Mackey, L.J. Smith and J.C. Wheeler for useful discussion. 

\bibliographystyle{mn2e}

\end{document}